\newif\ifCHANGES
\let\ifCHANGES\iftrue

\documentclass[
    aip,
    jcp,
    reprint,
    amssymb,
    superscriptaddress,
    floatfix,
]{revtex4-2}

\usepackage[utf8]{inputenc}

\usepackage{xparse}

\usepackage{amsmath}
\usepackage{amssymb}
\usepackage{mathtools}
\usepackage{bm}
\usepackage{chemformula}
\usepackage{physics}
\usepackage{siunitx}

\usepackage{graphicx}
\usepackage{xcolor}

\usepackage{todonotes}

\usepackage{acronym}
\usepackage{hyperref}

\usepackage{booktabs}

\definecolor{edits}{RGB}{220,0,0}
\definecolor{strike}{RGB}{150,50,50}
\ifCHANGES
    \usepackage[normalem]{ulem}
    
    \NewDocumentCommand\STRIKE{+m}{{\color{strike}\sout{#1}}}
\else
    
    \NewDocumentCommand\STRIKE{+m}{}
\fi

\NewDocumentCommand\eg{}{e.\,g.}
\NewDocumentCommand\ie{}{i.\,e.}
\NewDocumentCommand\cf{}{cf.}
\NewDocumentCommand\etal{s}{\emph{et~al\IfBooleanF{#1}{.}}}

\NewDocumentCommand\eqperiod{}{\,\text{.}}
\NewDocumentCommand\eqcomma{}{\,\text{,}}

\NewDocumentCommand\ee{}{\mathrm{e}}

\RenewDocumentCommand\vec{m}{\bm{#1}}

\DeclarePairedDelimiterX\set[1]{\{}{\}}
    {#1}

\DeclareSIUnit{\au}{a.u.}
\DeclareSIUnit{\aus}{a.u}

\NewDocumentCommand\kB{}{\ensuremath{k_\mathrm{B}}}
\NewDocumentCommand\rNC{}{\ensuremath{r_\mathrm{e}}}

\NewDocumentCommand\kMFPT{}{\ensuremath{k_\mathrm{MFPT}}}
\NewDocumentCommand\kMax{}{\ensuremath{k_\mathrm{max}}}

\begin{document}

\title{Mean first-passage times
for solvated LiCN isomerization
at intermediate to high temperatures}

\author{Micha M. Schleeh}
\altaffiliation{These authors contributed equally to this work.}

\author{Johannes Reiff}
\altaffiliation{These authors contributed equally to this work.}
\affiliation{%
    Institut für Theoretische Physik I,
    Universität Stuttgart,
    70550 Stuttgart, Germany
}

\author{Pablo L. Garc{\'\i}a-Müller}
\affiliation{%
    Departamento de Tecnología,
    Centro de Investigaciones Energéticas Medioambientales y Tecnológicas,
    Avda. Complutense 40, 28040 Madrid, Spain
}

\author{Rosa M. Benito}
\affiliation{%
    Grupo de Sistemas Complejos,
    Escuela Técnica Superior de Ingeniería Agronómica,
        Alimentaria y de Biosistemas,
    Universidad Politécnica de Madrid,
    28040 Madrid, Spain
}

\author{Florentino Borondo}
\affiliation{%
    Instituto de Ciencias Matemáticas (ICMAT),
    Cantoblanco, 28049 Madrid, Spain
}
\affiliation{%
    Departamento de Química,
    Universidad Autónoma de Madrid,
    Cantoblanco, 28049 Madrid, Spain
}

\author{Jörg Main}
\affiliation{%
    Institut für Theoretische Physik I,
    Universität Stuttgart,
    70550 Stuttgart, Germany
}

\author{Rigoberto Hernandez}
\email[Corresponding author: ]{r.hernandez@jhu.edu}
\affiliation{%
    Department of Chemistry,
    Johns Hopkins University,
    Baltimore, Maryland 21218, USA
}
\affiliation{%
    Departments of Chemical \& Biomolecular Engineering,
    and Materials Science and Engineering,
    Johns Hopkins University,
    Baltimore, Maryland 21218, USA
}

\date{\today}

\begin{abstract}
    The behavior of a particle in a solvent
    has been framed using stochastic dynamics since
    the early theory of Kramers.
    A particle in a chemical reaction reacts slower in a diluted solvent
    because of the lack of energy transfer via collisions.
    The flux-over-population reaction rate constant
    rises with increasing density
    before falling again for very dense solvents.
    This Kramers turnover is observed in this paper
    at intermediate and high temperatures in the backward reaction of the
    \ch{LiNC <=> LiCN} isomerization
    via Langevin dynamics and mean first-passage times (MFPTs).
    It is in good agreement with
    the Pollak--Grabert--Hänggi (PGH) reaction rates at lower temperatures.
    Furthermore, we find
    a square root behavior of the reaction rate at high temperatures
    and have made direct comparisons of the methods
    in the intermediate- and high- temperature regimes;
    all suggesting increased ranges in accuracy of both the
    PGH and MFPT approaches.
\end{abstract}

\maketitle


\acrodef{AAMD}{all-atom molecular dynamics}
\acrodef{DS}{dividing surface}
\acrodef{GLE}{generalized Langevin equation}
\acrodef{IRC}{intrinsic reaction coordinate}
\acrodef{LE}{Langevin equation}
\acrodef{MFPT}{mean first-passage time}
\acrodef{NHIM}{normally hyperbolic invariant manifold}
\acrodef{PES}{potential energy surface}
\acrodef{PGH}{Pollak--Grabert--Hänggi}
\acrodef{SBB}{Straub--Borkovec--Berne}
\acrodef{TST}{transition state theory}
\acrodef{TS}{transition state}


\section{Introduction}
\label{sec:introduction}

The \ac{MFPT} is a useful
estimate of an inverse reaction rate constant
(for simplicity called \emph{rate} in this paper)
whether it is determined numerically or
analytically.
\cite{talkner81,talkner97,rsh99,Redner2001a,vega02,hern02b,Park2003a}
For example, the inverse \ac{MFPT}
can be used to numerically obtain the Kramers rate\cite{rsh99}
through the direct observation of an ensemble of trajectories
of a given system.
Its use in resolving rates in chemical reactions has been demonstrated
in several systems including the
isomerization reaction of lithium cyanide (\ch{LiNC}/\ch{LiCN}).
The latter has received a lot of attention\cite{brocks1983ab,
borondo89a, borondo95a, borondo96a, borondo97a, Losada-2008, Prado2009,
borondo10, hern12e, hern14j, vergel2014geometrical, hern16c, hern20m}
in part because of the early introduction of a useful
model potential by Essers \etal*.\cite{essers1982scf}
Recently, for example,
some of us\cite{hern20m}
obtained the decay rate of trajectories
near the \ac{NHIM} for this reaction so as to demonstrate
the use of nonrecrossing dividing surfaces in
time-dependent \ac{TST}.\cite{hern21j}

In turn, the objective of this paper is to demonstrate
a stronger connection between rates obtained using \ac{MFPT}
to those from flux-over-population formulas.
Pollak, Grabert and Hänggi (\acs{PGH})\cite{pgh89}
employed the latter in constructing the theory
describing the Kramers turnover\cite{kram40}
in the rates from low to high friction.
The turnover was observed in the \ch{LiNC} reaction,\cite{hern08g}
and has subsequently been seen to correlate well
with the \ac{PGH} theory.\cite{hern12e,hern14j,hern16c}

Here, we obtain the \ac{MFPT} and the \ac{PGH} theory
explicitly for the backward reaction \ch{LiCN -> LiNC}.
Perhaps surprisingly,
we found that the \ac{MFPT} can be useful in describing
reactions of particles undergoing Brownian motion
at much higher temperature
than the barrier height.
Throughout this work, we use the terms low, intermediate, and high to qualify
our temperature range in reference to the barrier height.
Intermediate temperatures are on the order but lower than the barrier height,
whereas low temperatures are much lower than the barrier height by at least a
factor of 10 such that the system is mostly bound
and the reactions must be activated.
Notably, we benchmarked the
rates obtained from the \ac{MFPT} to
several alternative methods so as to confirm the relative accuracies
across a broad range of friction and temperature.

At high temperatures, the rate of escape and the corresponding rate constant
have to be carefully defined.
Here, a significant proportion of the initial conditions
are at initial energies above the barrier.
These are not trapped and simply escape across the barrier without return
(as long as an absorbing boundary condition is used on the other side).
These trajectories do not contribute to a steady-state flux
or the corresponding rate constant.
However, there is still a small population---perhaps fleetingly so---whose
energies are initially below the barrier and hence initially trapped.
They, of course, gain energy quickly through interactions with the bath
nearly instantaneously redistributing themselves
into yet another Boltzmann distribution.
Their rate of escape is the one that we calculate
and use to determine a rate constant.

The structure of the paper is as follows.
The model for the backward reaction \ch{LiCN -> LiNC} is summarized
in Sec.~\ref{sec:methods},
and connected to a solvent through the Langevin equation
in Sec.~\ref{sec:methods/langevin}.
The use of \ac{PGH} theory and \ac{MFPT} to obtain reaction
rates for this reaction
is presented in Secs.~\ref{sec:methods/pgh} and
\ref{sec:methods/mfpt}, respectively.
The \ac{MFPT} rates are presented in
Sec.~\ref{sec:results/mfpt},
and the nature of their turnover is addressed in
Sec.~\ref{sec:results/turnover}.
A comparison of these methods to \ac{AAMD}
for the prediction
of the maximum in the rates
in the challenging case of high temperatures
is presented in Sec.~\ref{sec:results/methodcmp}.


\section{Methods and materials}
\label{sec:methods}

\subsection{Non-driven isomerization of LiCN}
\label{sec:methods/licn}

\begin{figure}
    \includegraphics[width=\linewidth]{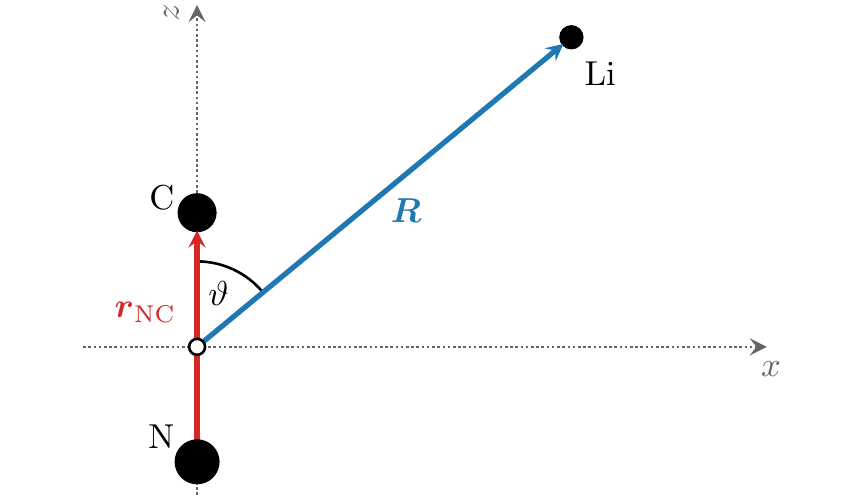}
    \caption{%
        Body-fixed Cartesian coordinate system
        of the \ch{LiNC <=> LiCN} isomerization reaction.
        The origin is located at the cyanide compound's center of mass.
        The Jacobi coordinates are defined in terms of
        the distance $R = \abs{\vec{R}}$ and
        the angle $\vartheta = \measuredangle(\vec{r}_{\ch{NC}}, \vec{R})$
        of the lithium atom relative to the origin.
        The position $\vec{r}_{\ch{NC}}$ of \ch{C} relative to \ch{N}
        and is assumed to be fixed because of the rigid \ch{N+C} bond.}
    \label{fig:licn_config}
\end{figure}

The isomerization reaction \ch{LiNC <=> LiCN}
involves the breaking and making of weak bonds
between a lithium cation and a cyanide anion.
The \ch{LiNC} is the stable conformation of the isomerization reaction,
and hence the backward reaction,
\begin{equation}
    \ch{Li-C+N -> C+N-Li}
    \eqcomma
\end{equation}
is exothermic.
Figure~\ref{fig:licn_config} shows the \ch{LiCN} molecule
in a Cartesian body-fixed frame.\cite{brocks1983ab}
Through the approximation of
separating the translation and rotation of the
center of mass of the whole molecule,
it is possible to reduce the equations of motion
to two internal variables.
These variables,
$R$ and $\vartheta$, are
known as the Jacobi coordinates:
$R = \abs{\bm{R}}$
is the distance between the \ch{NC} center of mass and the lithium atom,
and $\vartheta = \measuredangle(\vec{r}_{\ch{NC}}, \vec{R})$
is the angle between $\bm{R}$ and $\bm{r}_{\ch{NC}}$,
where $\abs{\bm{r}_{\ch{NC}}} = \rNC = \SI{2.186}{\bohr}$
and \si{\bohr} is the Bohr radius.
At $\vartheta = 0$, the potential is at a local minimum
that corresponds to the metastable conformation of
the \ch{LiCN} reactant in the backward reaction.
The global minimum of the potential lies at
$\vartheta = \pi$, and corresponds to
the stable conformation of the \ch{LiNC} product.

\begin{figure}
    \includegraphics[width=\linewidth]{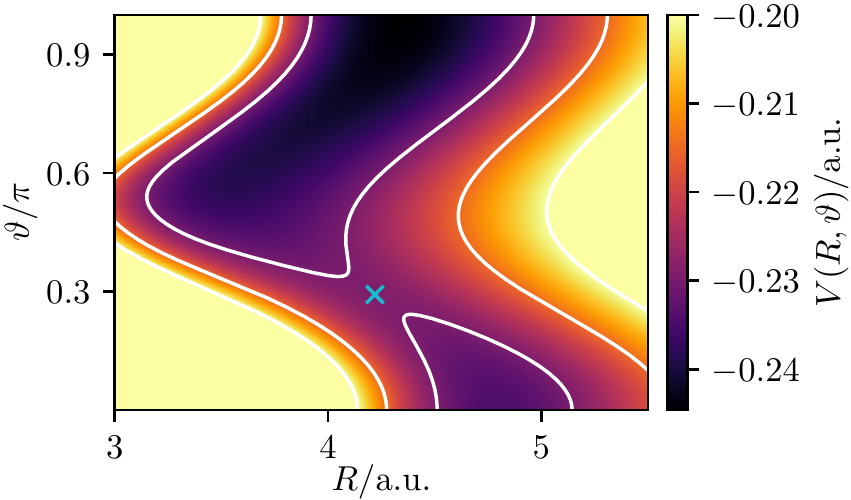}
    \caption{%
        Potential energy $V$ as a function of
        Jacobi coordinates $(R, \vartheta)$.
        The \acs{PES} shows a characteristic reaction channel
        with a saddle point at $V(4.2197, 0.2922 \pi) = \SI{-0.22886}{\au}$
        marked by the cyan cross.
        At $\vartheta = 0$ the isomer exists in its \ch{LiCN} configuration
        [$V(4.7947, 0) = \SI{-0.23421}{\au}$],
        whereas at $\vartheta = \pi$ it is in its \ch{LiNC} configuration
        [$V(4.3487, \pi) = \SI{-0.24461}{\au}$].}
    \label{fig:pot_saddle}
\end{figure}

The Hamiltonian can now be specified using
the Born--Oppenheimer potential energy surface\cite{essers1982scf}
and the usual approximations\cite{borondo89a, Losada-2008}
as
\begin{equation}
    \label{eq:licn_hamiltonian}
    \mathcal{H}
    = \frac{p_R^2}{2 \mu_1}
        + \frac{1}{R^2} \qty[\frac{p_\vartheta^2}{2 m^*(R)}]
        + V(R, \vartheta)
\end{equation}
containing the kinetic
and potential\cite{essers1982scf} $V(R, \vartheta)$
energy in $R$ and $\vartheta$
where the reduced masses are
$\mu_1 = [1 / m_{\ch{Li}} + 1 / (m_{\ch{C}} + m_{\ch{N}})]^{-1}$,
$\mu_2 = (1 / m_{\ch{C}} + 1 / m_{\ch{N}})^{-1}$, and
\begin{equation}
    m^*(R) = \qty(\frac{1}{\mu_1} + \frac{R^2}{\mu_2 \rNC^2})^{-1}
    \eqperiod
\end{equation}
We assume pure \ch{^{7}Li}, \ch{^{12}C}, and \ch{^{14}N} isotopes
for the atomic masses $m_{\ch{Li}}$, $m_{\ch{C}}$, and $m_{\ch{N}}$,
respectively.
The \ac{PES} in Jacobi coordinates
is shown in Fig.~\ref{fig:pot_saddle}.
The reaction saddle is marked with a cyan cross and
lies at  $R = \SI{4.2197}{\au}$ and $\vartheta = 0.2922 \pi$.

In this paper, we use atomic units (\si{\au})
for all of our figures and calculated values.
For example, the distances are given in terms of
the Bohr radius \si{\bohr},
the atomic time unit is given by $\hbar / \si{\hartree}$,
where \si{\hartree} is the Hartree energy,
and the mass is given in electron masses \si{\electronmass}.
In these units, we found high-temperature rates $k$ on the order of
\SI{e-4}{\au} (\cf\ Fig.~\ref{fig:methodcmp})
corresponding roughly to $k = \SI{4e12}{\per\s}$ in SI units.
The inverse of such a rate is comparable to
the time scale of the isomerization dynamics
reported for high excitation energies in Ref.~\onlinecite{Pellouchoud2015a},
which are also the dynamics accessed at high temperatures.


\subsection{Langevin implementation}
\label{sec:methods/langevin}

Solvent effects on the
\ch{LiNC <=> LiCN} isomerization reaction have previously been
addressed through the introduction of an
argon bath.\cite{hern08g,hern12e,hern14j}
The interaction with the bath can be reduced
to the Langevin equation though a mapping
to the characteristic friction and random noise.\cite{hern08g,hern12e,hern16c}
With the approximation
$(\dv*{t})(p_\vartheta / (m^* R)) \approx \dot{p}_\vartheta / (m^* R)$,
the equations of motion follow from
Hamiltonian~\eqref{eq:licn_hamiltonian} as
\begin{subequations}
    \label{eq:eom}
    \begin{align}
        \dot{\vartheta}
            &= \qty(\frac{1}{\mu_1 R^2} + \frac{1}{\mu_2 \rNC^2})
                p_\vartheta
        \eqcomma \\
        \label{eq:eom/p_theta}
        \dot{p}_\vartheta
            &= - \dv{V(R, \vartheta)}{\vartheta}
                - \gamma p_\vartheta
                + R \xi_\vartheta
        \eqcomma  \\
        \dot{R}
            &= \frac{p_R}{\mu_1}
        \eqcomma \\
        \dot{p}_R
            &= \frac{p_\vartheta^2}{\mu_1 R^3}
                - \dv{V(R, \vartheta)}{R}
                - \gamma p_R + \xi_R
        \eqcomma
    \end{align}
\end{subequations}
where the stochastic forces $\xi_\vartheta$ and $\xi_R$
satisfy the respective fluctuation-dissipation theorems\cite{kubo66}
\begin{subequations}
    \label{eq:fluctuationdissipation}
    \begin{align}
        \ev{\xi_{\vartheta, i}(t) \xi_{\vartheta, j}(t')}
            & = 6 \gamma \kB T m^* \delta_{i, j} \delta(t - t')
        \\ \qand*
        \ev{\xi_{R,i}(t) \xi_{R,j}(t')}
            & = 6  \gamma \kB T \mu_1 \delta_{i, j} \delta(t - t')
    \end{align}
\end{subequations}
for uniformly distributed noise.
The canonical momentum $p_\vartheta$ is an angular momentum.
It should therefore not be surprising that
the last term in Eq.~\eqref{eq:eom/p_theta} includes
a product with the radial coordinate $R$
as it leads to the correct units.
The random forces $\xi_i$ are generated at the beginning of the calculation
and use a
fixed $R$ ($= \SI{4.2196}{\au}$) at the barrier
with the reduced mass $m^*$ ($= \SI{2406}{\au}$).
This is a nontrivial approximation because the
reduced mass varies as much as \SI{25}{\percent}
across the positions in $R$, but it is consistent
with prior work and the error is least when
the trajectories are near the barrier.
A numerical consequence of this approximation is that the random forces
do not vary with $R$, thus allowing a significant
simplification in coding the equations of motion and
in implementing the theory.
Thus the price of this approximation is that the results may be affected in so
far as the effective temperature is renormalized.

The results have been calculated with a fourth order Langevin Runge--Kutta
keeping the random force constant during each time step.


\subsection{Pollak--Grabert--Hänggi theory}
\label{sec:methods/pgh}

The \ac{PGH} theory for activated processes
driven by Markovian forces has been seen to be very
effective in the low- and intermediate- temperature regimes
across the Kramers turnover of the rates with respect to friction.\cite{pgh89}
Among several examples,
\cite{Hanggi1986a, Straub1986a, Zwanzig1987a, berne88, pgh89, rmp90}
it was shown to be effective for characterizing the dynamics
across the potential model of \ac{SBB}.\cite{sbb85, sbb86}
The Langevin implementation is memoryless,
which is formally described by the friction kernel,
\begin{equation}
    \gamma(t, t') = \gamma_0 \delta(t - t')
    \eqperiod
\end{equation}
The \ac{SBB} approximation uses parabolic functions attached to each other
to create a single potential well
followed by a saddle as an inverse parabolic function,
and implemented in the generalized Langevin equation with
a memory friction.
It was used previously by some of us to model the minimum energy path of
the \ch{LiNC <=> LiCN} isomerization reaction.\cite{hern08g}
In the \ac{SBB} model, memory is introduced through a single
exponentially decaying term in the friction kernel,
\begin{equation}
    \gamma(t, t') = \alpha^{-1} \exp(-\frac{\abs{t - t'}}{\alpha \gamma})
    \eqperiod
\end{equation}
The propagation of particles inside this potential is strongly dependent on
the memory time scale $\tau = \alpha \gamma$ used in the friction kernel
and the form of the friction kernel $\gamma(t, t')$ itself.
For those relaxation processes that occur at times much longer than $\tau$,
the response looks ohmic as in the Langevin case.
However, the \ac{SBB} model now allows for dynamical responses from the
solvent that can compete with the dynamics in the system.
This leads to an effective friction which arises from the mean of the
modulate frictions from the previous times.
We found earlier\cite{hern16c} that in practice,
this led to an increase of observed rates by about a factor of \num{5}
when using the \ac{LE} rather than the \ac{SBB} model.


\subsection{Mean first-passage time rates}
\label{sec:methods/mfpt}

The first-passage time is the time a particle needs
to reach a certain region for the first time given some initial state.
In case of a reaction,
the first-passage time is defined as
the time the particle is propagated from a point in the reactant region
to a point on some characteristic surface at or beyond a \ac{DS}.
\cite{talkner81, Hill1989a, talkner97, rsh99, vega02, hern02b}
In reactive systems characterized by one-dimensional barriers,
the \ac{DS} reduces to a point.
It is naively taken to be the saddle point, but other choices are available,
just like for variational \ac{TST}, for example.
\cite{eyring35, wigner37, pech81, truh96, hern10a, peters14a, wiggins16, Ezra2018a}
The first-passage times for a series of trajectories
from different initial points in the reactive regime
experiencing varying thermal forces
vary stochastically.
Averaged together they lead to
the mean first-passage time $t_\mathrm{MFPT}$,
whose inverse is the rate of escape\cite{talkner87a, karplus88, rmp90, hern02b}
\begin{equation}
    \label{eq:mfpt_rate}
    \kMFPT = \frac{1}{t_\mathrm{MFPT}}
    \eqperiod
\end{equation}
In the limit of a harmonic barrier, \kMFPT\ has been seen to be
precisely equal to the \ac{TST} rate, and both
are equal to the correct Kramers rate $k$.\cite{rsh99}

Generally, the rate problem is treated exclusively in the activated regime.
Therein, the typical energies of the system
are characteristic of an average temperature that is well below
the energetic barrier.
The smooth turnover in the Kramers rates with friction
was resolved by Mel'nikov and Meshkov,\cite{mm86}
and Pollak, Grabert and H\"anggi (PGH).\cite{pgh89}
They found a mathematical expression for the rate
connecting the low-friction regime---where the rate increases with $\gamma$---%
and the high-friction regime---where the rate decreases with $1 / \gamma$.
\ac{PGH} imposes a rate-determining region (or \ac{DS} in
phase space) which requires temperatures to be low enough that
the reacting system is somehow thermalized.
Since our initial work\cite{hern08g,hern14j,hern16c}
demonstrating the applicability of the \ac{PGH} theory at surprisingly
higher temperatures, Pollak and coworkers have extended
it for temperatures near threshold.\cite{pollak16}

In the present problem, however, we must also consider
much higher temperatures
in which the reactive system usually accesses energies much
higher than the barrier along the reaction coordinate.
The process is consequently effectively barrierless,
and the rate problem reduces to the determination
of a steady-state flux for a given thermal molecular beam.
That is, the typical energies of those states
accessing and crossing the barrier at high temperatures%
---in the sense that they are much larger than the barrier---%
correspond to states that cross the
barrier freely (or ballistically) at their typical velocity.
As a consequence,
the rates reduce to a simple power law at high temperature,
\begin{equation}
    \label{eq:hightemperaturerates}
    k \propto \sqrt{\ev{v^2}} \propto \sqrt{T}
    \eqperiod
\end{equation}
This behavior is well known since the early work on
molecular beams.\cite{bernstein87, hasebook99}
Alternatively, we can recover Eq.~\eqref{eq:hightemperaturerates} from
the mean-squared displacement observed in small and long times
from the Langevin equation,\cite{Uhlenbeck1930a}
where the mean-squared displacement is given by
\begin{equation}
    \begin{split}
        \ev{\Delta r^2}
            &= \frac{2 N D}{\gamma} \qty(\gamma t - 1 + \ee^{-\gamma t})
            \\
            &\approx 2 N D \gamma t^2
            \\
            &\approx 2 N \kB T t^2
        \eqperiod
    \end{split}
\end{equation}
Thus the velocity dependence of Eq.~\eqref{eq:hightemperaturerates}
is again recovered when
\begin{equation}
    \label{eq:smalltimes}
    t \ll 1 / \gamma
\end{equation}
is satisfied,\cite{Uhlenbeck1930a}
and the reactants have a short \ac{MFPT} (and associated small $t$)
due to the fast barrierless crossing.


\section{Results}
\label{sec:results}


\subsection{Mean first-passage time rates of LiCN isomerization}
\label{sec:results/mfpt}

\begin{figure}
    \includegraphics[width=\linewidth]{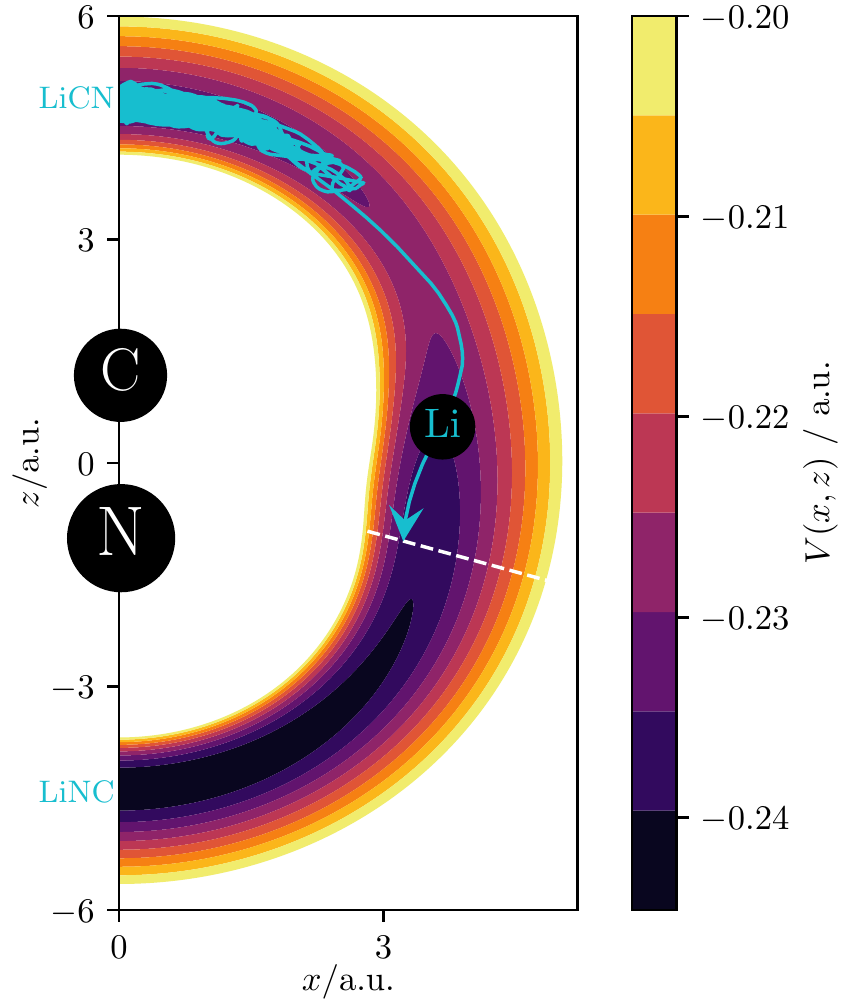}
    \caption{%
        Potential energy $V$
        as a function of body-fixed Cartesian coordinates $(x, z)$.
        The origin is located at the cyanide compound's center of mass.
        Filled circles illustrate
        the positions of the individual atoms in our model.
        The circles' radii are chosen proportional to the atomic masses.
        For lithium, an arbitrary position on
        an example trajectory at $T = \SI{300}{\K}$ is shown.
        The trajectory starts at
        the local minimum $x = \num{0}$, $z = \num{4.7947}$
        corresponding to the \ch{LiCN} configuration.
        It crosses the saddle located at $x = 3.3520$, $z = 2.5631$
        and ends at the absorbing boundary, indicated by the dashed line.
        The potential's global minimum $x = \num{0}$, $z = \num{-4.3487}$
        corresponds to the \ch{LiNC} isomer.}
    \label{fig:pot_trajectory}
\end{figure}

The \ac{PES} of \ch{LiNC <=> LiCN} (\cf\ Fig.~\ref{fig:pot_saddle})
is replotted in Fig.~\ref{fig:pot_trajectory}
in terms of the Cartesian coordinates, $x$ and $z$, of \ch{Li}
relative to the $C_{\infty \mathrm{v}}$ axis of \ch{CN}.
The minimum energy path of the potential
nearly follows a semi-circle with a radius of $R \approx \SI{4.5}{\aus}$.
The minimum energy of the
\ch{LiCN} molecule ($E_{\mathrm{min}, \ch{LiCN}} = \SI{-0.23421}{\au}$)
is not as deep as that of \ch{LiNC}
($E_{\mathrm{min}, \ch{LiNC}} = \SI{-0.24461}{\au}$).
Consequently, a trajectory starting at the \ch{LiCN} state
has to overcome a smaller barrier height of
$E^\ddag = \SI{0.00535}{\au}$ (corresponds to $E^\ddag / \kB = \SI{1690}{\K}$)
than one starting from the \ch{LiNC} state.
At room temperature ($T = \SI{300}{\K}$), for example,
the barrier of this backward reaction
is thus low enough that such trajectories
are activated to above threshold energy frequently enough
that they can be observed numerically.
This is also true even for the forward reaction despite its higher barrier.

To ensure that selected trajectories
are properly identified as reactive, they must
first cross the saddle
and then reach the product side without turning around.
This condition is satisfied using an
absorbing boundary\cite{rmp90, Kappler2018a}
defined by an angle of $\varphi = 0.6 \pi$,
and shown as the dotted white line
in the Cartesian \ac{PES} plot of Fig.~\ref{fig:pot_trajectory}.
Indeed, if the trajectory reaches this line,
then it generally has enough momentum in the direction of the product side
to make it very unlikely for it to turn around
and climb back over the saddle to the reactant side.
Use of absorbing boundary conditions along space and/or energy to
address low to mid friction also has significant precedent
in the literature.\cite{mm86, pgh89, hern01d}
In the present case focusing on high temperatures,
we did not add an energy condition to the absorbing boundary
because the spacial constraint is sufficient to remove
the high-energy trajectories once they have reached the
product basin (\cf\ Fig.~\ref{fig:pot_trajectory})
before their long-time return to reactants.
Its use enables us to better compare the \ac{MFPT} calculations to
the \ac{AAMD} results obtained earlier in Ref.~\onlinecite{hern12e}
where direct rates were determined based on
the flux through a \ac{DS} along the reaction coordinate.\cite{hern08g}

The representative trajectory shown in Fig.~\ref{fig:pot_trajectory}
is only one of \num{1500} propagated trajectories
from the ensemble
used to calculate the \ac{MFPT} rate \kMFPT.
Trajectories of this thermal ensemble are initialized
at a specific temperature $T$ and located near the reactive well.
For simplicity in the current implementation,
they are all placed at the \ch{LiCN} minimum at
$R = \SI{4.7947}{\au}$ and $\vartheta = 0$,
which in Cartesian coordinates corresponds to
$x = \SI{0}{\au}$ and $z = \SI{4.7947}{\aus}$.
Their velocity is sampled from
a Maxwell--Boltzmann distribution with temperature $T$.
The assumption here is that
the redistribution to a quasi-bound distribution in space
is fast compared to the escape times when the system is trapped.
Meanwhile, in the barrierless cases, there would be no such trapping
and we are merely considering the steady-state flux
from the injection of particles at this origin.
This case could presumably be accessible
through increasingly exquisite spectroscopic methods
such as \ac{TS} spectroscopy.
\cite{Polanyi1995a, Wenthold1996a, Neumark1996a, Hamm2009a}

At higher bath temperatures,
the distributions in energy---total, potential and kinetic---of
the initial reactant ensemble necessarily shift to higher values.
Consequently, a larger percentage of the states in the reactant ensemble
will possess the required energy to surmount the barrier at any given time.
More of these states escape over the barrier and are removed from the simulation.
The remaining and decreasing population of reactive states is re-thermalized
according to the bath temperature,
and thereby continues to contribute to the steady-state rate.
As the temperature increases further,
this activation is faster
leading to increasingly larger kinetic energies in the trajectories
contributing to the steady-state rate.
This, in turn, causes smaller \acp{MFPT} and higher rates.
Specifically,
ensembles thermalized at some temperature $T$
and propagated at a certain friction value $\gamma$
lead to rates that depend
not on the fast inertial trajectories but rather those
that endure several traversals of the reaction region%
---at energies below activation threshold---%
before escaping.


\subsection{Kramers turnover}
\label{sec:results/turnover}

\begin{figure}
    \includegraphics[width=\linewidth]{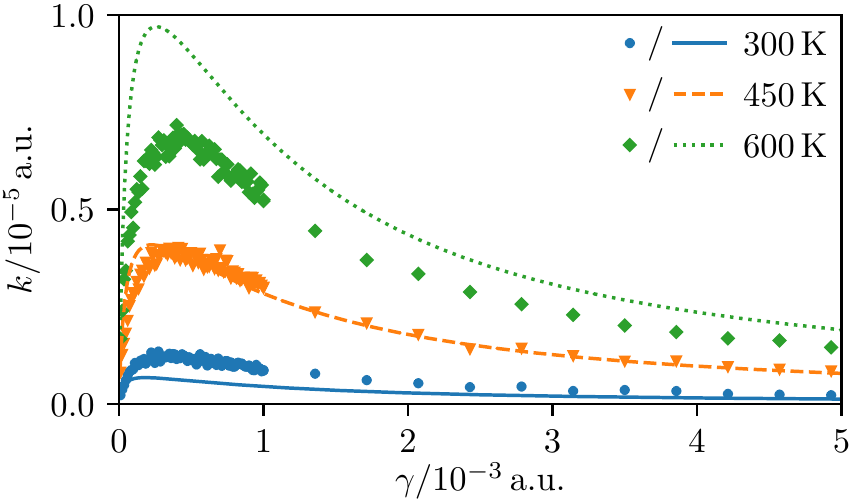}
    \caption{%
        Mean first-passage time rates $\kMFPT / 5$
        of the \ch{LiCN -> LiNC} backward reaction
        as a function of friction $\gamma$
        at temperatures $T = \SI{300}{\K}$ (blue circles),
        $T = \SI{450}{\K}$ (orange triangles),
        and $T = \SI{600}{\K}$ (green diamonds).
        For comparison, the corresponding \acs{PGH} rates
        obtained for a \acs{GLE} model%
        ---with friction kernel specified by
        a bath parameter $\alpha = \SI{1.5625}{\au}$
        and decay time $\tau$ ($= \alpha \gamma$)---%
        are shown as
        solid blue, dashed orange, and dotted green lines, respectively.
        The reported rates $\kMFPT / 5$ are obtained
        from the \acs{MFPT} rates
        for the \acs{LE} model (with ohmic friction),
        and scaled by a factor of \num{5}
        to make the shapes of the Kramers turnovers comparable.}
    \label{fig:kramersturnovers}
\end{figure}

We now report the \ac{MFPT} rates \kMFPT\
across the friction domain at several
intermediate temperatures in Fig.~\ref{fig:kramersturnovers}.
As before,\cite{hern08g, hern12e, hern16c}
we observe a clear Kramers turnover,
and the shape of \kMFPT\ is
in very good agreement with the
corresponding \ac{PGH} rate formula.
The rate maxima are always around $\gamma \approx \SI{4e-4}{\aus}$.

To obtain the rates in Fig.~\ref{fig:kramersturnovers},
different ensembles at $T = \SI{300}{\K}$,
$T = \SI{450}{\K}$, and $T = \SI{600}{\K}$ are thermalized
on each friction value $\gamma$.
Each mean rate \kMFPT\ is calculated using
\num{1500} trajectories propagated by the Langevin equation.
At low friction ($\SI{1e-5}{\au} \ll \gamma \ll \SI{2e-4}{\au}$),
we find the expected linear increasing behavior of the rates.
Similarly, at high friction,
the rates decrease strongly with $1 / \gamma$.
These two limits have been known since the work of Kramers.
In combination, they give rise to the eponymous Kramers turnover.
In absolute values, however, the \ac{MFPT} rates overestimate
the expected rates by a factor of \num{5}.
This is unfortunate, but not unexpected, because 
the \ac{MFPT} rates---just like those from flux over population\cite{rsh99}---%
are known to be approximate due to several factors.
In the present implementation, these \acp{MFPT} are 
uniformly shorter, leading to faster rates than measured 
through the other approaches, 
presumably because 
there is an acceleration from the initial distribution that 
is not being dissipated by the bath.

The rates calculated using trajectories propagated via the \ac{LE}
are also overestimated by about a factor of \num{5}
compared to the \ac{SBB} model and \ac{AAMD} calculations.
Specifically, the Kramers turnover rate maxima \kMax\
in the \ac{LE} and \ac{GLE} from Ref.~\onlinecite{hern16c}
are in agreement only if the rates obtained
for the reactive flux calculations for the
\ac{LE} were to be divided by a factor of \num{5}.
Meanwhile, the \ac{GLE} rates obtained using \ac{PGH} theory
for the near ohmic limit (at very small $\alpha$) 
do not suffer from this overestimation.
This suggests that the \ac{LE} rates calculated using reactive flux
suffer from numerical error not present in the \ac{GLE} dynamics. 
We conjecture, though have not proven, that this discrepancy arises
from a slower energetic redistribution among the quasi-bound states
allowing for a faster escape of the initial higher energy 
configurations.

\begin{figure}
    \includegraphics[width=\linewidth]{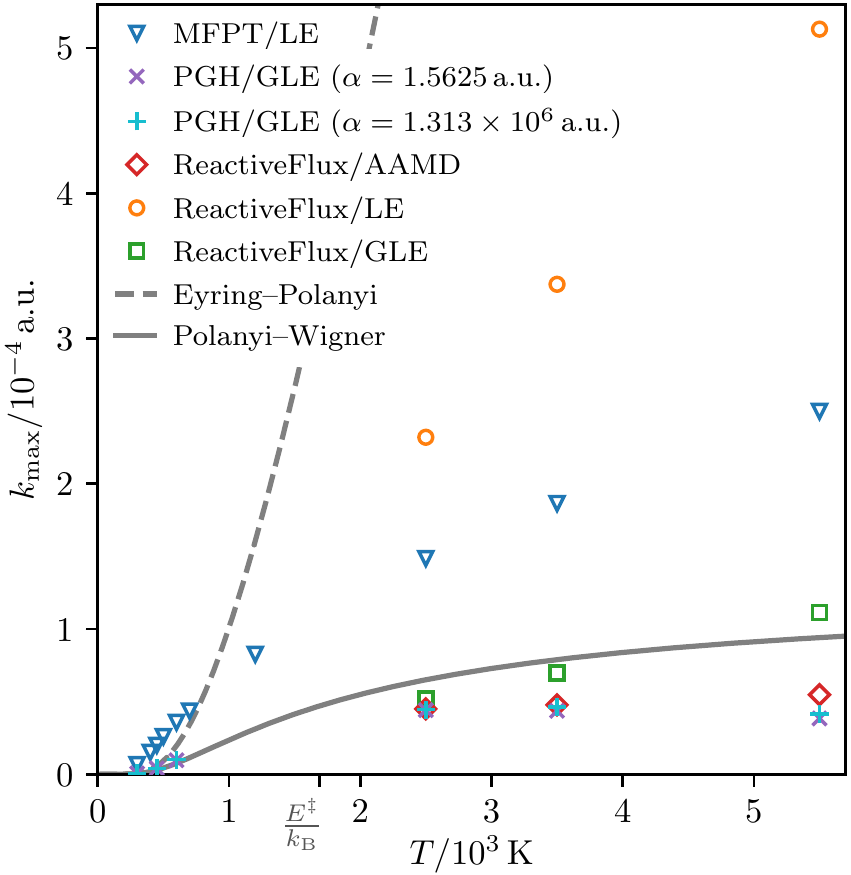}
    \caption{%
        Kramers turnover rate maxima \kMax\ as functions of temperature $T$.
        The rates \kMFPT\ (blue triangles),
        calculated with an \acs{LE} friction kernel,
        are shown in the low- and high-temperature regimes,
        which are separated by the effective temperature of the barrier
        $E^\ddag / \kB = \SI{1690}{\K}$.
        The reported \acs{PGH} rates,
        marked with purple crosses and cyan pluses,
        are obtained for the \acs{GLE} model with bath parameters
        $\alpha = \SI{1.5625}{\au}$ (corresponding to short-time memory)
        and $\alpha = \SI{1.313e6}{\au}$ (corresponding to long-time memory),
        respectively.
        The \acs{AAMD}- (red diamonds), \acs{LE}- (orange circles),
        and \acs{GLE}-based (green squares)
        reactive-flux rates\cite{hern12e, hern16c}
        are calculated via the flux-over-population method,
        and are available only in the high-temperature regime.
        For reference, rate calculations based on
        two \acs{TST} formulas (\cf\ Appendix~\ref{sec:simplerate})
        are shown with dashed and solid gray lines.}
    \label{fig:methodcmp}
\end{figure}

To confirm the effect of friction in the Kramers turnover of \kMFPT\
seen in the \ac{PGH} theory in Fig.~\ref{fig:kramersturnovers},
the bath parameter $\alpha$ in the \ac{PGH} friction kernel
is set to a small value $\alpha = \SI{1.5625}{\aus}$.
The maximal rates for fixed $\alpha$ in this case are still found
at friction values $\gamma_0$ in a similar range
and hence this case ($\alpha = \SI{1.5625}{\au}$)
indeed leads to a short memory time scale of $\tau \approx \SI{6.25e-4}{\aus}$.
This places it in the memoryless dynamics regime
characteristic of the \ac{LE}, as noted above.
Meanwhile, the \ac{PGH} rates obtained
at the longer memory time $\tau \approx \SI{5.25e2}{\au}$%
---corresponding to a larger $\alpha$---%
as was observed in the \ac{AAMD} simulations
still exhibits similar maximal rates
in Fig.~\ref{fig:methodcmp}.
At intermediate temperatures, for example,
the memory time scale is still much lower than the \ac{MFPT}
($t_\mathrm{MFPT} \gtrsim \SI{1e4}{\au}$),
and thus the system remains in a nearly memoryless regime.
As the \ac{PGH} theory%
---which is also known to be
in good agreement with \ac{TST}\cite{rmp90}
at the appropriate limits---%
describes the low- and intermediate- temperature regime well,
it is not surprising that the \ac{MFPT} rates are effective
as long as we ignore a temperature-independent uniform factor.


\subsection{High-temperature regime}
\label{sec:results/methodcmp}

The \ac{MFPT} rates can also be used
to describe the high-temperature regime,
where the typical energies are well
above the barrier height $E^\ddag / \kB = \SI{1690}{\K}$.
Therein the Kramers turnover maxima in the \ac{MFPT} rates
are compared to the corresponding reactive-flux rates\cite{hern12e}
in Fig.~\ref{fig:methodcmp}.
Specifically, we compare against
\ac{LE}-, \ac{GLE}-, and \ac{AAMD}-based calculations,
where we consider \ac{AAMD} to yield the most accurate results.
Notably, the latter includes cavity reorganization effects
which may not be fully described within
the \ac{LE}-based reactive-flux and \ac{MFPT} approaches.

Figure~\ref{fig:methodcmp} also includes \ac{TST} benchmark calculations
based on the Eyring--Polanyi (\cf\ Appendix~\ref{sec:simplerate/eyring})
and the Polanyi--Wigner (\cf\ Appendix~\ref{sec:simplerate/polanyi})
rate expressions.
While the Eyring--Polanyi rate describes the reaction rate reasonably well
for $\kB T \ll E^\ddag$, as is well known,%
\cite{truh83,hynes85b,sbb88,rmp90,tucker91,truh96,chandler00,truh2000,Berne2016}
it quickly deviates from the other results
as the temperature is increased.
Indeed, the usual rule of thumb is that there is a breakdown
in rate theories---because they begin to violate the separation
between reactant and transition state regions---above this threshold.
Thus the result in the all-atom \ch{LiCN} dynamics in
an argon solvent seen in Refs.~\onlinecite{hern08g, hern12e}
was surprising because it suggested the existence of an observable
rate at such high temperatures, and their turnover with apparent friction
obeyed the Kramers turnover seen through the lens of \ac{PGH} theory.
Here, we see that part of the origin of the breakdown in the
Eyring--Polanyi theory is that it suggests an infinite rate which
is not possible when the particles move at velocities distributed around
a Boltzmann distribution for a given $\kB T$.
Instead, at high enough temperatures,
the reactants move at apparent rates
that are proportional to their average velocity because at such energies
the barrier is not visible.
The results in Fig.~\ref{fig:methodcmp}
demonstrate that the rates, as also captured by Polanyi--Wigner, move
smoothly from the well-known low-temperature regime through intermediate
temperatures in which the barrier plays a role, albeit a decreasing one
with increasing temperature.

Thus at increasing temperatures, the reaction is no longer limited by the
activation process that is inherent in the Eyring--Polanyi formula
and rises much faster leading to a very large upper bound of the rate.
On the other hand, the Polanyi--Wigner formalism appears to capture the
slowing down in the rates due to the decreasing quasi-bound population
in the reactive basin with increasing temperature.
Specifically, the prefactor in the Polanyi--Wigner rate is no longer
temperature dependent but rather determined by
the vibrational frequency $\omega_0 / 2 \pi \approx \SI{3.2e12}{\per\s}$
of the reactant well and, therefore, leads to a flattening in the rate
with increasing high temperatures.
Nevertheless, the interaction with the solvent manifests itself
by way of the decreased rates through the intermediate temperature
regime.

The comparison between the maxima of the \ac{PGH} rates
and the \ac{AAMD} rates for higher temperatures in Ref.~\onlinecite{hern12e}
is also shown in Fig.~\ref{fig:methodcmp}.
The agreement shows that the \ac{PGH} theory
can also be applied to high temperatures.
In this work, we further found that the \ac{PGH} rates
are able to describe the intermediate- and high- temperature regime,
both with the memory time scale appropriate to the system (cyan pluses)
and through the approximate Langevin approach (purple crosses).
This also justifies the use of the \kMFPT\ obtained from the
\ac{LE} friction kernel at the high temperatures.

The \ac{AAMD}, \ac{GLE} reactive-flux, and \ac{PGH} results
reported in Fig.~\ref{fig:methodcmp}
are in good agreement at high temperatures.
As already observed in Ref.~\onlinecite{hern16c},
the reactive-flux results based on the \ac{LE} grossly overestimate the rates
by a factor of \numrange{4.5}{4.8} and \numrange{5.1}{9.3} compared to
those seen in the \ac{GLE} and \ac{AAMD} reactive-flux calculations,
respectively.
The \ac{MFPT} rates for the \ac{LE} are lower than those from the
reactive-flux calculations on the \ac{LE}
but not enough to match the other methods.
Specifically, \kMFPT\ overestimates by a factor between
\num{2.9} at $T = \SI{2500}{\K}$ and \num{2.2} at $T = \SI{5500}{\K}$
compared to the \ac{GLE} reactive flux.
The overestimation factor is larger
when compared to \ac{AAMD} calculations,
monotonically rising
from \num{3.3} at $T = \SI{2500}{\K}$ to \num{4.6} at $T = \SI{5500}{\K}$.
There is, however, no such simple trend
when considering the whole temperature range via comparison with \ac{PGH} rates.
Instead,
the overestimation factor is particularly high at very low temperatures
(\num{9.7} at $T = \SI{300}{\K}$),
quickly drops until it reaches a local minimum at $T = \SI{2500}{\K}$,
and finally rises again slowly with temperature.

Compared to the \ac{LE} reactive-flux approach,
the \ac{MFPT} appears to better capture
the escape of the quasi-trapped initial states at high temperatures.
These low-probability states
contribute to the steady-state flux obtained from the \ac{AAMD},
\ac{GLE} reactive flux and \ac{PGH} leading to an even lower rate.
The disagreement between the latter and those obtained using the \ac{LE}
is surprising because the \ac{PGH} results for a \ac{GLE}
with a very small $\alpha$
is formally the exact answer for an \ac{LE} and should
be in direct agreement with the reactive flux obtained on the \ac{LE}.
Thus, both the \ac{MFPT} and reactive-flux calculations of the \ac{LE}
may be suffering from numerical errors due to the fact
that the needed contributing trajectories are rare.
The better agreement obtained with the \ac{MFPT}
may reflect the fact that it does not suffer
as much from the constraints related to the selection of the initial states from
the quasi-bound population.

Despite some of the loss in accuracy
arising from the strong localization of the initial
distribution of states,
the advantage of the \ac{MFPT} rates compared to \ac{AAMD} rates is that
they can be calculated for the whole temperature regime in less time.
Furthermore, a square root function shape is observable for \kMFPT,
as described in Sec.~\ref{sec:methods/mfpt}.
This shape arises because the rate is known to depend
on the square root of temperature in this regime.

The difference between \kMFPT\
and reactive-flux-based approaches in Fig.~\ref{fig:methodcmp}
is a consequence of the assumptions inherent in each method.
For the reactive-flux rate,
an extended ensemble is initialized at the minimum of \ch{LiCN}
and the rates are obtained by the flux-over-population method
across numerically integrated trajectories.
In the \ac{GLE}, the rates are obtained through a direct
(and analytic) analysis of the reactive ensemble across the barrier.
In principle, long-time returns to the barrier
may be absent from the latter leading
to an overestimate in the \acl{TST} calculation.
Such long-time sojourns are long only in the sense that they are longer
than the effective reactive timescales but may be short in an absolute sense
because the temperatures (and the kinetic energies) are large.
A related consequence of the high-temperature regime is that the
kinetic energy tends to be dominated by equipartition leading the
effective average rate across the barrier to simply be described
by the barrierless average velocity.
The ensuing square root behavior is seen correctly in the
\ac{MFPT} rates in Fig.~\ref{fig:methodcmp}
at high temperatures
whereas the \ac{PGH} rates are unfortunately linear.


\section{Conclusion and outlook}
\label{sec:conclusion}

In this work, we have calculated the rates of
the \ch{LiCN -> LiNC} backward reaction using \aclp{MFPT}.
We benchmarked them relative to the results of earlier
\ac{AAMD} calculations and
the \ac{PGH} rates across several
temperatures and friction regimes.
The \ac{MFPT} rates are effective in describing
the Kramers turnover in the intermediate temperature regime
as shown by comparisons to the \ac{PGH} rates
calculated with a short memory time scale.
At very high temperatures when the reactants motion effectively
becomes ballistic, the \ac{MFPT} approach was seen to
correctly capture the square-root dependence of the rates
on temperature.

At low temperatures, there are several examples
of near-perfect unadjusted agreement between methods like
the reactive flux and \ac{GLE}-based theories like
Grote--Hynes.\cite{Grote-1980, pollak86b, Peters2017a}
See, for example, reports of the rates in
cyclohexane interconversion,\cite{Peters04a}
\ch{NaCl} dissociation,\cite{Ciccotti1990a, Rey1992a}
S\textsubscript{N}2 reactions,\cite{bergsma87, hynes87, Gertner1989a}
and even some enzymes.\cite{Roca2006a, RuizPernia2008a, Kanaan2010a}
However, for the \ch{LiCN} isomerization reaction at intermediate
and high temperatures, we found, as before,\cite{hern16c} that
methods based on Langevin dynamics overestimate the rate
roughly by a factor of \num{5}
compared to \ac{AAMD} or \ac{GLE} simulations.
The program used for our paper was written independently of
the one used in Ref.~\onlinecite{hern16c}.
It therefore seems unlikely that
this factor is merely caused by a bug in some code.
Instead, this suggests a physical origin which we conjecture here,
as done earlier in Ref.~\onlinecite{hern16c},
that it originates in otherwise unaccounted
mean field dissipation in the determination of the stochastic trajectories.
Thus the \ac{MFPT} approach pursued in this paper was found to
provide no worse agreement than other methods while being
relatively simple to implement in systems with arbitrary dimensionality.
This overestimation nevertheless poses a challenge for future work.

Furthermore, comparisons of the computed rate constants to
experiment would clarify which theoretical approach
is most suitable to describe the \ch{LiCN} isomerization reaction.
The experimental measurement of these rates
especially in the high-temperature regime,
also poses a challenge for future work.
Specifically, for \ch{LiCN}, it may be difficult to
construct an initial distribution of states
localized at the reactant
well that is weakly solvated by a thermalized argon bath.
Perhaps, spectroscopic scalpels such as those offered by
\ac{TS} spectroscopy%
\cite{Polanyi1995a, Wenthold1996a, Neumark1996a, Hamm2009a}
could be employed,
and we look forward to seeing such advances.

In addition, the \ac{MFPT} approach provides a deeper view
of the structure of the \ch{LiCN} isomerization reaction.
The footprint (or projection) of the ensemble of trajectories
that contribute to the rate onto the domain of reactant coordinates
appears to spread across the \ac{IRC}.\cite{Fukui1970, hern20f}
Though not addressed quantitatively here, we conjecture that this
spread can be used to characterize the manifold (or tube) that envelopes
the possible reaction pathway,\cite{hern10a}
but now constructed organically without
the assumptions required of the \ac{IRC}.

This work also opens at least two new possible directions to address rates in
chemical reactions and other activated systems:
The first lies in the need for more direct comparison between
\ac{NHIM} rates\cite{hern20m} with \ac{MFPT} rates.
The second lies in the use of these approaches to obtain
the rates for a time-dependent \ch{LiCN} system
driven by an electrical field so as to fully demonstrate
the computational advantages of this approach in chemical reactions generally.
Finally, reproducing the calculations of the
\ac{MFPT} on the Langevin equation with
those for the general Langevin equation (with the
observed memory time of the solvent)
could also confirm the generality of our finding.


\appendix
\section{LiCN potential surface}
\label{sec:pot}

The \ch{LiNC <=> LiCN} \ac{PES} according to Ref.~\onlinecite{essers1982scf}
consists of two parts:
a damped long-range energy plus a short-range energy.

\begin{table}
    \caption{%
        Expectation values $\ev{Q_{L, 0}}$ of
        the \ch{CN-} multipole moments
        used in Eq.~\eqref{eq:pot/electrostatic}
        and induction energy coefficients $C_{l_1, l_2, L}$
        used in Eq.~\eqref{eq:pot/induction}.
        Originally published in Refs.~\onlinecite{essers1982scf}
        and (partially)~\onlinecite{wormer1981abinitio}.
        The bold value $C_{2, 1, 3}$ differs from the original publication.}
    \label{tab:licn_longrange}
    \centering
    \begin{tabular}{lrrrrrrr}
        \toprule
        L & $\ev{Q_{L, 0}}$ & $C_{1, 1, L}$ & $C_{2, 1, L}$ & $C_{2, 2, L}$ & $C_{3, 1, L}$ & $C_{3, 2, L}$ & $C_{3, 3, L}$ \\
        \midrule
        0 & $-1.00$ & $-10.53$ & & $-57.49$ & & & $-458.2$ \\
        1 & $-0.2151$ & & $-10.31$ & & & $-101.45$ & \\
        2 & $-3.414$ & $-3.17$ & & $-35.71$ & $-35.56$ & & $-353.7$ \\
        3 & $-3.819$ & & $\mathbf{1.866}$ & & & $-37.62$ & \\
        4 & $-15.84$ & & & $5.23$ & $5.95$ & & $-112.6$ \\
        5 & $-14.29$ & & & & & $-14.23$ & \\
        6 & $-43.82$ & & & & & & $-108.3$ \\
        \bottomrule
    \end{tabular}
\end{table}

The long-range part is composed of the electrostatic energy
\begin{equation}
    \label{eq:pot/electrostatic}
    E_\mathrm{el}(R, \vartheta)
    = \sum_{L = 0}^\infty R^{-L - 1} P_L(\cos \vartheta) \ev{Q_{L, 0}}
\end{equation}
and the induction energy
\begin{multline}
    \label{eq:pot/induction}
    E_\mathrm{ind}(R, \vartheta) =
    \\
    \sum_{l_1, l_2 = 0}^\infty R^{-l_1 - l_2 - 2}
        \sum_{L = \abs{l_1 - l_2}}^{l_1 + l_2}
            P_L(\cos \vartheta) C_{l_1, l_2, L}
    \eqperiod
\end{multline}
Here, $P_L$ is the Legendre polynomial of order $L$,
$\ev{Q_{L, 0}}$ denotes the expectation value of
the order-$L$ \ch{CN-} multipole moment,
and $C_{l_1, l_2, L}$ are the induction coefficients.
Numerical values for $\ev{Q_{L, 0}}$ and $C_{l_1, l_2, L}$
are given in Table~\ref{tab:licn_longrange}.
The damping is represented by
\begin{equation}
    F(R) = 1 - \exp[-a \qty(R - R_0)^2]
\end{equation}
with fit parameters $a = \SI{1.5156}{\au}$ and $R_0 = \SI{1.9008}{\aus}$.

\begin{table}
    \caption{%
        Parameters $A_L$, $B_L$, and $C_L$ found for
        the analytical expression of the short-range interaction
        in Eq.~\eqref{eq:pot/srfit}
        as fitted to the potential
        originally published in Ref.~\onlinecite{essers1982scf}.
        The bold value $C_2$ differs from that in the original publication,
        but all of the other values are the same.}
    \label{tab:licn_shortrange}
    \centering
    \begin{tabular}{
        S[table-format=1] S[table-format=-2.5]
        S[table-format=-1.5] S[table-format=-1.6]
    }
        \toprule
        {$L$} & {$A_L$} & {$B_L$} & {$C_L$} \\
        \midrule
        0 & -1.38321 &  0.14001 &  0.207892 \\
        1 & -2.95791 &  1.47977 & -0.011613 \\
        2 & -4.74203 &  1.81199 & \textbf{-0.017181} \\
        3 & -1.88853 &  1.28750 &  0.027728 \\
        4 & -4.41433 &  2.32297 & -0.070693 \\
        5 & -4.02565 &  2.77538 & -0.137720 \\
        6 & -5.84259 &  3.48085 & -0.186331 \\
        7 & -2.61681 &  2.65559 & -0.005882 \\
        8 & -6.34466 &  4.34498 & -0.152914 \\
        9 & 15.2023  & -6.54925 &  1.302568 \\
        \bottomrule
    \end{tabular}
\end{table}

The short-range term can be written as
\begin{equation}
    E_\mathrm{SR}(R, \vartheta)
    = \sum_{L = 0}^\infty D_L(R) P_L(\cos \vartheta)
    \eqcomma
\end{equation}
where $D_L(R)$ has been fitted to the analytical form
\begin{equation}
    \label{eq:pot/srfit}
    D_L(R) = \exp(-A_L - B_L R - C_L R^2)
    \eqperiod
\end{equation}
Numerical values for the fit parameters $A_L$, $B_L$, and $C_L$
are given in Table~\ref{tab:licn_shortrange}.

Combining long and short-range energies, the final \ac{PES} reads
\begin{multline}
    V(R, \vartheta) =
    \\
    \qty[E_\mathrm{el}(R, \vartheta) + E_\mathrm{ind}(R, \vartheta)] F(R)
        + E_\mathrm{SR}(R, \vartheta)
    \eqperiod
\end{multline}

\begin{figure}
    \includegraphics[width=\linewidth]{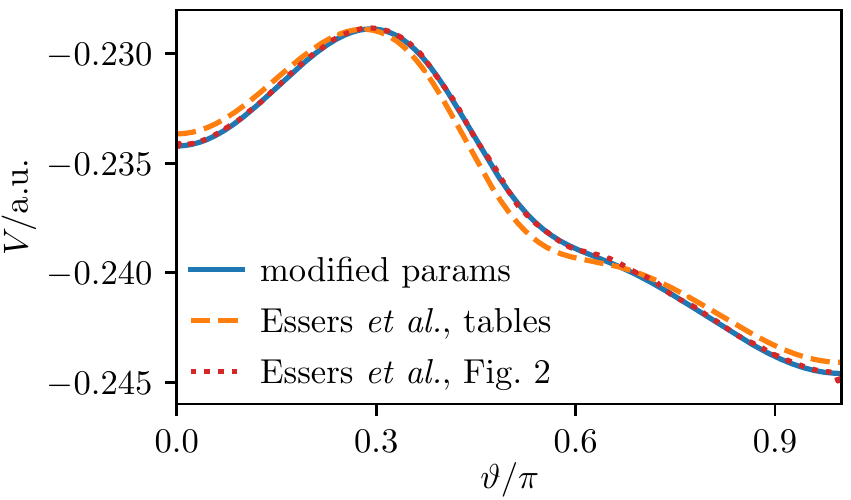}
    \caption{%
        Potential energy $V$ as a function of angle $\vartheta$
        on the minimum energy path of
        the \ch{LiNC <=> LiCN} isomerization reaction.
        The dotted line was extracted from
        Fig.~2 of Ref.~\onlinecite{essers1982scf}.
        This curve differs from what can be obtained
        using the parameters published in the same article (dashed line).
        Modifying the parameters as detailed in Sec.~\ref{sec:pot}
        yields much better agreement, as illustrated by the solid line.}
    \label{fig:mepcomp}
\end{figure}

We noticed a discrepancy while comparing the minimum energy path calculated
using the parameters published in Ref.~\onlinecite{essers1982scf}
with Fig.~2 from Ref.~\onlinecite{essers1982scf}.
As shown in Fig.~\ref{fig:mepcomp}, the two curves differ visibly.
Two parameters in the original source code\cite{TennysonLiCNProcedure}
differ significantly from the originally published values,
possibly due to errors introduced during the paper's production process.
The correct values%
---shown in bold in
Tables~\ref{tab:licn_longrange} and~\ref{tab:licn_shortrange}---%
yield a much better agreement with Fig.~2 from Ref.~\onlinecite{essers1982scf}.

An implementation of the potential in the Python programming language
using the same parameters as
the original source code\cite{TennysonLiCNProcedure}
can be found on GitHub.\cite{reiff21licn}


\section{Approximative rate formulas}
\label{sec:simplerate}

Most approximative rate formulas follow the
Arrhenius form\cite{Arrhenius1889, rmp90}
\begin{equation}
    \label{eq:simplerate/arrhenius}
    k(T) = \nu(T) \exp(-\frac{E^\ddag}{\kB T})
\end{equation}
where $\nu$ is a possibly temperature-dependent prefactor,
$E^\ddag$ is the reaction's barrier height or activation energy,
\kB\ is the Boltzmann constant,
and $T$ is the temperature.
In the following we present two important \ac{TST} variants of this equation.


\subsection{Polanyi--Wigner rate}
\label{sec:simplerate/polanyi}

One of the earliest results of \ac{TST}
is the unimolecular rate equation
derived, amongst others, by Polanyi and Wigner in 1928.\cite{Polanyi1928a}
It follows the Arrhenius rate law~\eqref{eq:simplerate/arrhenius}
with the pre-exponential factor $\nu(T)$ given by
the vibrational frequency $\omega_0 / 2 \pi$ of the reactant well\cite{rmp90}
\begin{equation}
    \label{eq:simplerate/polanyi}
    k(T) = \frac{\omega_0}{2 \pi} \exp(-\frac{E^\ddag}{\kB T})
    \eqcomma
\end{equation}
where $E^\ddag$ is the internal energy difference between the barrier
and the reactant of the isolated system.
This equation has been derived in various contexts.
It can, \eg, be recovered from the underdamped regime
$\omega^\ddag \kB T / E^\ddag \ll \gamma \ll \omega^\ddag$
of Kramers's medium-to-high-viscosity rate\cite{kram40, mm86}
\begin{equation}
    k(T) =
        \qty[\sqrt{\qty(\frac{\gamma}{2 \omega^\ddag})^2 + 1}
		    - \frac{\gamma}{2 \omega^\ddag}]
        \frac{\omega_0}{2 \pi} \exp(-\frac{E^\ddag}{\kB T})
    \eqcomma
\end{equation}
where $\gamma$ is the friction
and $\omega^\ddag$ is the inverse barrier frequency.
Equation~\eqref{eq:simplerate/polanyi} can therefore be seen as
a \ac{TST} rate which is
an upper bound for the rate at the turnover in Kramers's original theory
for a \textit{solvated} reaction.\cite{rmp90}


\subsection{Eyring--Polanyi rate}
\label{sec:simplerate/eyring}

The usual (or modern) form for the classical \ac{TST} rate equation
is given by the Eyring--Polanyi equation\cite{eyring35, eyring35b, Laidler1983}
\begin{equation}
    k(T) = \frac{\kappa \kB T}{h} \exp(-\frac{\Delta G^\ddag}{\kB T})
\end{equation}
with $\kappa = 1$,
where $\kappa$ is the transmission coefficient,
$h$ is Planck's constant,
and $\Delta G^\ddag$ is the Gibbs energy of activation
in the context of a solvent.

The Gibbs energy of activation can be approximately determined
from the enthalpy of activation $\Delta H^\ddag$ via
\begin{equation}
    \Delta G^\ddag = \Delta H^\ddag - T \Delta S^\ddag
    \eqcomma
\end{equation}
where $\Delta S^\ddag$ is the entropy of activation.
In turn, the enthalpy of activation for an unimolecular gas-phase reaction
can be written as
\begin{equation}
    \Delta H^\ddag = E^\ddag - \kB T
    \eqperiod
\end{equation}
The energy of activation $E^\ddag$
and the entropy of activation $\Delta S^\ddag$
can finally be determined from the minimum energy path
and the potential of mean force of the reaction\cite{hern08g, hern12e}
by equating the latter with the Gibbs energy.

The transmission coefficient $\kappa$ in the Eyring--Polanyi equation
describes the fraction of states
that cross the \ac{DS} between reactants and products at most once,
\ie, those that do not recross.
This quantity cannot be determined from straightforward statistical mechanics
and is therefore of great interest\cite{eyring35b} in the general case.
In \ac{TST}, it is assumed to be approximately one.\cite{rmp90}
This approximation is valid if the temperature is not too high
or if the friction is sufficiently strong.


\begin{acknowledgments}
    Useful discussions with Thomas Bartsch are gratefully acknowledged.
    The German portion of this collaborative work was partially supported
    by the Deutsche Forschungsgemeinschaft (DFG) through Grant No.~MA1639/14-1.
    The Spanish portion was supported by
    the Spanish Ministry of Science, Innovation and Universities (MICIU)
    under Contract No.~PGC2018-093854-B-I00.
    The US portion was partially supported
    by the National Science Foundation (NSF) through Grant No.~CHE 1700749.
    This collaboration has also benefited from support
    by the European Union's Horizon 2020 Research and Innovation Program
    under the Marie Skłodowska-Curie Grant Agreement No.~734557.
\end{acknowledgments}


\section*{Data Availability}

The data that support the findings of this study are available
from the corresponding author upon reasonable request.


\section*{References}

\bibliography{paper-q35}

\begin{thebibliography}{84}%
\makeatletter
\providecommand \@ifxundefined [1]{%
 \@ifx{#1\undefined}
}%
\providecommand \@ifnum [1]{%
 \ifnum #1\expandafter \@firstoftwo
 \else \expandafter \@secondoftwo
 \fi
}%
\providecommand \@ifx [1]{%
 \ifx #1\expandafter \@firstoftwo
 \else \expandafter \@secondoftwo
 \fi
}%
\providecommand \natexlab [1]{#1}%
\providecommand \enquote  [1]{``#1''}%
\providecommand \bibnamefont  [1]{#1}%
\providecommand \bibfnamefont [1]{#1}%
\providecommand \citenamefont [1]{#1}%
\providecommand \href@noop [0]{\@secondoftwo}%
\providecommand \href [0]{\begingroup \@sanitize@url \@href}%
\providecommand \@href[1]{\@@startlink{#1}\@@href}%
\providecommand \@@href[1]{\endgroup#1\@@endlink}%
\providecommand \@sanitize@url [0]{\catcode `\\12\catcode `\$12\catcode
  `\&12\catcode `\#12\catcode `\^12\catcode `\_12\catcode `\%12\relax}%
\providecommand \@@startlink[1]{}%
\providecommand \@@endlink[0]{}%
\providecommand \url  [0]{\begingroup\@sanitize@url \@url }%
\providecommand \@url [1]{\endgroup\@href {#1}{\urlprefix }}%
\providecommand \urlprefix  [0]{URL }%
\providecommand \Eprint [0]{\href }%
\providecommand \doibase [0]{https://doi.org/}%
\providecommand \selectlanguage [0]{\@gobble}%
\providecommand \bibinfo  [0]{\@secondoftwo}%
\providecommand \bibfield  [0]{\@secondoftwo}%
\providecommand \translation [1]{[#1]}%
\providecommand \BibitemOpen [0]{}%
\providecommand \bibitemStop [0]{}%
\providecommand \bibitemNoStop [0]{.\EOS\space}%
\providecommand \EOS [0]{\spacefactor3000\relax}%
\providecommand \BibitemShut  [1]{\csname bibitem#1\endcsname}%
\let\auto@bib@innerbib\@empty
\bibitem [{\citenamefont {H{\"a}nggi}\ and\ \citenamefont
  {Talkner}(1981)}]{talkner81}%
  \BibitemOpen
  \bibfield  {author} {\bibinfo {author} {\bibfnamefont {P.}~\bibnamefont
  {H{\"a}nggi}}\ and\ \bibinfo {author} {\bibfnamefont {P.}~\bibnamefont
  {Talkner}},\ }\bibfield  {title} {\enquote {\bibinfo {title} {Non-markov
  processes: the problem of the mean first passage time},}\ }\href
  {https://doi.org/10.1007/BF01294279} {\bibfield  {journal} {\bibinfo
  {journal} {Z. Physik B}\ }\textbf {\bibinfo {volume} {45}},\ \bibinfo {pages}
  {79--83} (\bibinfo {year} {1981})}\BibitemShut {NoStop}%
\bibitem [{\citenamefont {M{\"u}ller}, \citenamefont {Talkner},\ and\
  \citenamefont {Reimann}(1997)}]{talkner97}%
  \BibitemOpen
  \bibfield  {author} {\bibinfo {author} {\bibfnamefont {R.}~\bibnamefont
  {M{\"u}ller}}, \bibinfo {author} {\bibfnamefont {P.}~\bibnamefont
  {Talkner}},\ and\ \bibinfo {author} {\bibfnamefont {P.}~\bibnamefont
  {Reimann}},\ }\bibfield  {title} {\enquote {\bibinfo {title} {Rates and mean
  first passage times},}\ }\href
  {https://doi.org/10.1016/S0378-4371(97)00390-7} {\bibfield  {journal}
  {\bibinfo  {journal} {Physica A}\ }\textbf {\bibinfo {volume} {247}},\
  \bibinfo {pages} {338--356} (\bibinfo {year} {1997})}\BibitemShut {NoStop}%
\bibitem [{\citenamefont {Reimann}, \citenamefont {Schmid},\ and\ \citenamefont
  {H{\"a}nggi}(1999)}]{rsh99}%
  \BibitemOpen
  \bibfield  {author} {\bibinfo {author} {\bibfnamefont {P.}~\bibnamefont
  {Reimann}}, \bibinfo {author} {\bibfnamefont {G.~J.}\ \bibnamefont
  {Schmid}},\ and\ \bibinfo {author} {\bibfnamefont {P.}~\bibnamefont
  {H{\"a}nggi}},\ }\bibfield  {title} {\enquote {\bibinfo {title} {Universal
  equivalence of mean first-passage time and {Kramers} rate},}\ }\href
  {https://doi.org/10.1103/PhysRevE.60.R1} {\bibfield  {journal} {\bibinfo
  {journal} {Phys. Rev. E}\ }\textbf {\bibinfo {volume} {60}},\ \bibinfo
  {pages} {R1} (\bibinfo {year} {1999})}\BibitemShut {NoStop}%
\bibitem [{\citenamefont {Redner}(2001)}]{Redner2001a}%
  \BibitemOpen
  \bibfield  {author} {\bibinfo {author} {\bibfnamefont {S.}~\bibnamefont
  {Redner}},\ }\href {https://doi.org/10.1017/cbo9780511606014} {\emph
  {\bibinfo {title} {A Guide to First-Passage Processes}}}\ (\bibinfo
  {publisher} {Cambridge University Press},\ \bibinfo {year}
  {2001})\BibitemShut {NoStop}%
\bibitem [{\citenamefont {Vega}, \citenamefont {Guantes},\ and\ \citenamefont
  {Miret-Art{\'e}s}(2002)}]{vega02}%
  \BibitemOpen
  \bibfield  {author} {\bibinfo {author} {\bibfnamefont {J.~L.}\ \bibnamefont
  {Vega}}, \bibinfo {author} {\bibfnamefont {R.}~\bibnamefont {Guantes}},\ and\
  \bibinfo {author} {\bibfnamefont {S.}~\bibnamefont {Miret-Art{\'e}s}},\
  }\bibfield  {title} {\enquote {\bibinfo {title} {Mean first passage time and
  the {Kramers} turnover theory in activated atom-surface diffusion},}\ }\href
  {https://doi.org/10.1039/B204462E} {\bibfield  {journal} {\bibinfo  {journal}
  {Phys. Chem. Chem. Phys.}\ }\textbf {\bibinfo {volume} {4}},\ \bibinfo
  {pages} {4985} (\bibinfo {year} {2002})}\BibitemShut {NoStop}%
\bibitem [{\citenamefont {Shepherd}\ and\ \citenamefont
  {Hernandez}(2002)}]{hern02b}%
  \BibitemOpen
  \bibfield  {author} {\bibinfo {author} {\bibfnamefont {T.~D.}\ \bibnamefont
  {Shepherd}}\ and\ \bibinfo {author} {\bibfnamefont {R.}~\bibnamefont
  {Hernandez}},\ }\bibfield  {title} {\enquote {\bibinfo {title} {An optimized
  mean first passage time approach for obtaining rates in activated
  processes},}\ }\href {https://doi.org/10.1063/1.1516590} {\bibfield
  {journal} {\bibinfo  {journal} {J. Chem. Phys.}\ }\textbf {\bibinfo {volume}
  {117}},\ \bibinfo {pages} {9227--9233} (\bibinfo {year} {2002})}\BibitemShut
  {NoStop}%
\bibitem [{\citenamefont {Park}\ \emph {et~al.}(2003)\citenamefont {Park},
  \citenamefont {Sener}, \citenamefont {Lu},\ and\ \citenamefont
  {Schulten}}]{Park2003a}%
  \BibitemOpen
  \bibfield  {author} {\bibinfo {author} {\bibfnamefont {S.}~\bibnamefont
  {Park}}, \bibinfo {author} {\bibfnamefont {M.~K.}\ \bibnamefont {Sener}},
  \bibinfo {author} {\bibfnamefont {D.}~\bibnamefont {Lu}},\ and\ \bibinfo
  {author} {\bibfnamefont {K.}~\bibnamefont {Schulten}},\ }\bibfield  {title}
  {\enquote {\bibinfo {title} {Reaction paths based on mean first-passage
  times},}\ }\href {https://doi.org/10.1063/1.1570396} {\bibfield  {journal}
  {\bibinfo  {journal} {J. Chem. Phys.}\ }\textbf {\bibinfo {volume} {119}},\
  \bibinfo {pages} {1313--1319} (\bibinfo {year} {2003})}\BibitemShut {NoStop}%
\bibitem [{\citenamefont {Brocks}\ and\ \citenamefont
  {Tennyson}(1983)}]{brocks1983ab}%
  \BibitemOpen
  \bibfield  {author} {\bibinfo {author} {\bibfnamefont {G.}~\bibnamefont
  {Brocks}}\ and\ \bibinfo {author} {\bibfnamefont {J.}~\bibnamefont
  {Tennyson}},\ }\bibfield  {title} {\enquote {\bibinfo {title} {Ab initio
  rovibrational spectrum of {LiNC} and {LiCN}},}\ }\href
  {https://doi.org/10.1016/0022-2852(83)90312-0} {\bibfield  {journal}
  {\bibinfo  {journal} {J. Mol. Spectrosc.}\ }\textbf {\bibinfo {volume}
  {99}},\ \bibinfo {pages} {263--278} (\bibinfo {year} {1983})}\BibitemShut
  {NoStop}%
\bibitem [{\citenamefont {Benito}\ \emph {et~al.}(1989)\citenamefont {Benito},
  \citenamefont {Borondo}, \citenamefont {Kim}, \citenamefont {Sumpter},\ and\
  \citenamefont {Ezra}}]{borondo89a}%
  \BibitemOpen
  \bibfield  {author} {\bibinfo {author} {\bibfnamefont {R.~M.}\ \bibnamefont
  {Benito}}, \bibinfo {author} {\bibfnamefont {F.}~\bibnamefont {Borondo}},
  \bibinfo {author} {\bibfnamefont {J.-H.}\ \bibnamefont {Kim}}, \bibinfo
  {author} {\bibfnamefont {B.~G.}\ \bibnamefont {Sumpter}},\ and\ \bibinfo
  {author} {\bibfnamefont {G.~S.}\ \bibnamefont {Ezra}},\ }\bibfield  {title}
  {\enquote {\bibinfo {title} {Comparison of classical and quantum phase space
  structure of nonrigid molecules, {LiCN}},}\ }\href
  {https://doi.org/10.1016/S0009-2614(89)87032-0} {\bibfield  {journal}
  {\bibinfo  {journal} {Chem. Phys. Lett.}\ }\textbf {\bibinfo {volume}
  {161}},\ \bibinfo {pages} {60--66} (\bibinfo {year} {1989})}\BibitemShut
  {NoStop}%
\bibitem [{\citenamefont {Borondo}, \citenamefont {Zembekov},\ and\
  \citenamefont {Benito}(1995)}]{borondo95a}%
  \BibitemOpen
  \bibfield  {author} {\bibinfo {author} {\bibfnamefont {F.}~\bibnamefont
  {Borondo}}, \bibinfo {author} {\bibfnamefont {A.~A.}\ \bibnamefont
  {Zembekov}},\ and\ \bibinfo {author} {\bibfnamefont {R.~M.}\ \bibnamefont
  {Benito}},\ }\bibfield  {title} {\enquote {\bibinfo {title} {Quantum
  manifestations of saddle-node bifurcations},}\ }\href
  {https://doi.org/10.1016/0009-2614(95)01147-X} {\bibfield  {journal}
  {\bibinfo  {journal} {Chem. Phys. Lett.}\ }\textbf {\bibinfo {volume}
  {246}},\ \bibinfo {pages} {421} (\bibinfo {year} {1995})}\BibitemShut
  {NoStop}%
\bibitem [{\citenamefont {Borondo}, \citenamefont {Zembekov},\ and\
  \citenamefont {Benito}(1996)}]{borondo96a}%
  \BibitemOpen
  \bibfield  {author} {\bibinfo {author} {\bibfnamefont {F.}~\bibnamefont
  {Borondo}}, \bibinfo {author} {\bibfnamefont {A.~A.}\ \bibnamefont
  {Zembekov}},\ and\ \bibinfo {author} {\bibfnamefont {R.~M.}\ \bibnamefont
  {Benito}},\ }\bibfield  {title} {\enquote {\bibinfo {title} {Saddle‐node
  bifurcations in the linc/licn molecular system: Classical aspects and quantum
  manifestations},}\ }\href {https://doi.org/2048/10.1063/1.472351} {\bibfield
  {journal} {\bibinfo  {journal} {J. Chem. Phys.}\ }\textbf {\bibinfo {volume}
  {105}},\ \bibinfo {pages} {5068} (\bibinfo {year} {1996})}\BibitemShut
  {NoStop}%
\bibitem [{\citenamefont {Zembekov}\ \emph {et~al.}(1997)\citenamefont
  {Zembekov}, \citenamefont {Borondo}, \citenamefont {Zembekov},\ and\
  \citenamefont {Benito}}]{borondo97a}%
  \BibitemOpen
  \bibfield  {author} {\bibinfo {author} {\bibfnamefont {A.~A.}\ \bibnamefont
  {Zembekov}}, \bibinfo {author} {\bibfnamefont {F.}~\bibnamefont {Borondo}},
  \bibinfo {author} {\bibnamefont {Zembekov}},\ and\ \bibinfo {author}
  {\bibfnamefont {R.~M.}\ \bibnamefont {Benito}},\ }\bibfield  {title}
  {\enquote {\bibinfo {title} {Semiclassical quantization of fragmented tori:
  Application to saddle-node states of linc/licn},}\ }\href
  {https://doi.org/2048/10.1063/1.475147} {\bibfield  {journal} {\bibinfo
  {journal} {J. Chem. Phys.}\ }\textbf {\bibinfo {volume} {107}},\ \bibinfo
  {pages} {7934} (\bibinfo {year} {1997})}\BibitemShut {NoStop}%
\bibitem [{\citenamefont {Losada}, \citenamefont {Benito},\ and\ \citenamefont
  {Borondo}(2008)}]{Losada-2008}%
  \BibitemOpen
  \bibfield  {author} {\bibinfo {author} {\bibfnamefont {J.~C.}\ \bibnamefont
  {Losada}}, \bibinfo {author} {\bibfnamefont {R.~M.}\ \bibnamefont {Benito}},\
  and\ \bibinfo {author} {\bibfnamefont {F.}~\bibnamefont {Borondo}},\
  }\bibfield  {title} {\enquote {\bibinfo {title} {Frequency map analysis of
  the {3D} vibrational dynamics of the {LiCN}/{LiNC} molecular system},}\
  }\href {https://doi.org/10.1140/epjst/e2008-00862-0} {\bibfield  {journal}
  {\bibinfo  {journal} {Eur. Phys. J. Spec. Top.}\ }\textbf {\bibinfo {volume}
  {165}},\ \bibinfo {pages} {183--193} (\bibinfo {year} {2008})}\BibitemShut
  {NoStop}%
\bibitem [{\citenamefont {Prado}\ \emph {et~al.}(2009)\citenamefont {Prado},
  \citenamefont {Vergini}, \citenamefont {Benito},\ and\ \citenamefont
  {Borondo}}]{Prado2009}%
  \BibitemOpen
  \bibfield  {author} {\bibinfo {author} {\bibfnamefont {S.~D.}\ \bibnamefont
  {Prado}}, \bibinfo {author} {\bibfnamefont {E.~G.}\ \bibnamefont {Vergini}},
  \bibinfo {author} {\bibfnamefont {R.~M.}\ \bibnamefont {Benito}},\ and\
  \bibinfo {author} {\bibfnamefont {F.}~\bibnamefont {Borondo}},\ }\bibfield
  {title} {\enquote {\bibinfo {title} {Superscars in the {LiNC}-{LiCN}
  isomerization reaction},}\ }\href
  {https://doi.org/10.1209/0295-5075/88/40003} {\bibfield  {journal} {\bibinfo
  {journal} {Europhys. Lett.}\ }\textbf {\bibinfo {volume} {88}},\ \bibinfo
  {pages} {40003} (\bibinfo {year} {2009})}\BibitemShut {NoStop}%
\bibitem [{\citenamefont {Murgida}\ \emph {et~al.}(2010)\citenamefont
  {Murgida}, \citenamefont {Wisniacki}, \citenamefont {Tamborenea},\ and\
  \citenamefont {Borondo}}]{borondo10}%
  \BibitemOpen
  \bibfield  {author} {\bibinfo {author} {\bibfnamefont {G.~E.}\ \bibnamefont
  {Murgida}}, \bibinfo {author} {\bibfnamefont {D.~A.}\ \bibnamefont
  {Wisniacki}}, \bibinfo {author} {\bibfnamefont {P.~I.}\ \bibnamefont
  {Tamborenea}},\ and\ \bibinfo {author} {\bibfnamefont {F.}~\bibnamefont
  {Borondo}},\ }\bibfield  {title} {\enquote {\bibinfo {title} {Control of
  chemical reactions using external electric fields: {The} case of the
  {LiNC$\rightleftharpoons$LiCN} isomerization},}\ }\href
  {https://doi.org/10.1016/j.cplett.2010.07.057} {\bibfield  {journal}
  {\bibinfo  {journal} {Chem. Phys. Lett.}\ }\textbf {\bibinfo {volume}
  {496}},\ \bibinfo {pages} {356--361} (\bibinfo {year} {2010})}\BibitemShut
  {NoStop}%
\bibitem [{\citenamefont {{Garc{\'\i}a-M\"uller}}\ \emph
  {et~al.}(2012)\citenamefont {{Garc{\'\i}a-M\"uller}}, \citenamefont
  {Hernandez}, \citenamefont {Benito},\ and\ \citenamefont
  {Borondo}}]{hern12e}%
  \BibitemOpen
  \bibfield  {author} {\bibinfo {author} {\bibfnamefont {P.~L.}\ \bibnamefont
  {{Garc{\'\i}a-M\"uller}}}, \bibinfo {author} {\bibfnamefont {R.}~\bibnamefont
  {Hernandez}}, \bibinfo {author} {\bibfnamefont {R.~M.}\ \bibnamefont
  {Benito}},\ and\ \bibinfo {author} {\bibfnamefont {F.}~\bibnamefont
  {Borondo}},\ }\bibfield  {title} {\enquote {\bibinfo {title} {Detailed study
  of the direct numerical observation of the {Kramers} turnover in the
  {LiNC$\rightleftharpoons$LiCN} isomerization rate},}\ }\href
  {https://doi.org/10.1063/1.4766257} {\bibfield  {journal} {\bibinfo
  {journal} {J. Chem. Phys.}\ }\textbf {\bibinfo {volume} {137}},\ \bibinfo
  {pages} {204301} (\bibinfo {year} {2012})}\BibitemShut {NoStop}%
\bibitem [{\citenamefont {{Garc{\'\i}a-M\"uller}}\ \emph
  {et~al.}(2014)\citenamefont {{Garc{\'\i}a-M\"uller}}, \citenamefont
  {Hernandez}, \citenamefont {Benito},\ and\ \citenamefont
  {Borondo}}]{hern14j}%
  \BibitemOpen
  \bibfield  {author} {\bibinfo {author} {\bibfnamefont {P.~L.}\ \bibnamefont
  {{Garc{\'\i}a-M\"uller}}}, \bibinfo {author} {\bibfnamefont {R.}~\bibnamefont
  {Hernandez}}, \bibinfo {author} {\bibfnamefont {R.~M.}\ \bibnamefont
  {Benito}},\ and\ \bibinfo {author} {\bibfnamefont {F.}~\bibnamefont
  {Borondo}},\ }\bibfield  {title} {\enquote {\bibinfo {title} {{The role of
  the CN vibration in the activated dynamics of {LiNC$\rightleftharpoons$LiCN}
  isomerization in an argon solvent at high temperatures}},}\ }\href
  {https://doi.org/10.1063/1.4892921} {\bibfield  {journal} {\bibinfo
  {journal} {J. Chem. Phys.}\ }\textbf {\bibinfo {volume} {141}},\ \bibinfo
  {pages} {074312} (\bibinfo {year} {2014})}\BibitemShut {NoStop}%
\bibitem [{\citenamefont {Vergel}\ \emph {et~al.}(2014)\citenamefont {Vergel},
  \citenamefont {Benito}, \citenamefont {Losada},\ and\ \citenamefont
  {Borondo}}]{vergel2014geometrical}%
  \BibitemOpen
  \bibfield  {author} {\bibinfo {author} {\bibfnamefont {A.}~\bibnamefont
  {Vergel}}, \bibinfo {author} {\bibfnamefont {R.~M.}\ \bibnamefont {Benito}},
  \bibinfo {author} {\bibfnamefont {J.~C.}\ \bibnamefont {Losada}},\ and\
  \bibinfo {author} {\bibfnamefont {F.}~\bibnamefont {Borondo}},\ }\bibfield
  {title} {\enquote {\bibinfo {title} {Geometrical analysis of the {LiCN}
  vibrational dynamics: A stability geometrical indicator},}\ }\href
  {https://doi.org/10.1103/PhysRevE.89.022901} {\bibfield  {journal} {\bibinfo
  {journal} {Phys. Rev. E}\ }\textbf {\bibinfo {volume} {89}},\ \bibinfo
  {pages} {022901} (\bibinfo {year} {2014})}\BibitemShut {NoStop}%
\bibitem [{\citenamefont {Junginger}\ \emph {et~al.}(2016)\citenamefont
  {Junginger}, \citenamefont {{Garc{\'\i}a-M\"uller}}, \citenamefont {Borondo},
  \citenamefont {Benito},\ and\ \citenamefont {Hernandez}}]{hern16c}%
  \BibitemOpen
  \bibfield  {author} {\bibinfo {author} {\bibfnamefont {A.}~\bibnamefont
  {Junginger}}, \bibinfo {author} {\bibfnamefont {P.~L.}\ \bibnamefont
  {{Garc{\'\i}a-M\"uller}}}, \bibinfo {author} {\bibfnamefont {F.}~\bibnamefont
  {Borondo}}, \bibinfo {author} {\bibfnamefont {R.~M.}\ \bibnamefont
  {Benito}},\ and\ \bibinfo {author} {\bibfnamefont {R.}~\bibnamefont
  {Hernandez}},\ }\bibfield  {title} {\enquote {\bibinfo {title} {Solvated
  molecular dynamics of {LiCN} isomerization: All-atom argon solvent versus a
  generalized {Langevin} bath},}\ }\href {https://doi.org/10.1063/1.4939480}
  {\bibfield  {journal} {\bibinfo  {journal} {J. Chem. Phys.}\ }\textbf
  {\bibinfo {volume} {144}},\ \bibinfo {pages} {024104} (\bibinfo {year}
  {2016})}\BibitemShut {NoStop}%
\bibitem [{\citenamefont {Feldmaier}\ \emph {et~al.}(2020)\citenamefont
  {Feldmaier}, \citenamefont {Reiff}, \citenamefont {Benito}, \citenamefont
  {Borondo}, \citenamefont {Main},\ and\ \citenamefont {Hernandez}}]{hern20m}%
  \BibitemOpen
  \bibfield  {author} {\bibinfo {author} {\bibfnamefont {M.}~\bibnamefont
  {Feldmaier}}, \bibinfo {author} {\bibfnamefont {J.}~\bibnamefont {Reiff}},
  \bibinfo {author} {\bibfnamefont {R.~M.}\ \bibnamefont {Benito}}, \bibinfo
  {author} {\bibfnamefont {F.}~\bibnamefont {Borondo}}, \bibinfo {author}
  {\bibfnamefont {J.}~\bibnamefont {Main}},\ and\ \bibinfo {author}
  {\bibfnamefont {R.}~\bibnamefont {Hernandez}},\ }\bibfield  {title} {\enquote
  {\bibinfo {title} {Influence of external driving on decays in the geometry of
  the {LiCN} isomerization},}\ }\href {https://doi.org/10.1063/5.0015509}
  {\bibfield  {journal} {\bibinfo  {journal} {J. Chem. Phys.}\ }\textbf
  {\bibinfo {volume} {153}},\ \bibinfo {pages} {084115} (\bibinfo {year}
  {2020})}\BibitemShut {NoStop}%
\bibitem [{\citenamefont {Essers}, \citenamefont {Tennyson},\ and\
  \citenamefont {Wormer}(1982)}]{essers1982scf}%
  \BibitemOpen
  \bibfield  {author} {\bibinfo {author} {\bibfnamefont {R.}~\bibnamefont
  {Essers}}, \bibinfo {author} {\bibfnamefont {J.}~\bibnamefont {Tennyson}},\
  and\ \bibinfo {author} {\bibfnamefont {P.~E.~S.}\ \bibnamefont {Wormer}},\
  }\bibfield  {title} {\enquote {\bibinfo {title} {An {SCF} potential energy
  surface for lithium cyanide},}\ }\href
  {https://doi.org/10.1016/0009-2614(82)80046-8} {\bibfield  {journal}
  {\bibinfo  {journal} {Chem. Phys. Lett.}\ }\textbf {\bibinfo {volume} {89}},\
  \bibinfo {pages} {223--227} (\bibinfo {year} {1982})}\BibitemShut {NoStop}%
\bibitem [{\citenamefont {Nagahata}, \citenamefont {Hernandez},\ and\
  \citenamefont {Komatsuzaki}(2021)}]{hern21j}%
  \BibitemOpen
  \bibfield  {author} {\bibinfo {author} {\bibfnamefont {Y.}~\bibnamefont
  {Nagahata}}, \bibinfo {author} {\bibfnamefont {R.}~\bibnamefont
  {Hernandez}},\ and\ \bibinfo {author} {\bibfnamefont {T.}~\bibnamefont
  {Komatsuzaki}},\ }\bibfield  {title} {\enquote {\bibinfo {title} {Phase space
  geometry of isolated to condensed chemical reactions},}\ }\href
  {https://doi.org/10.1063/5.0059618} {\bibfield  {journal} {\bibinfo
  {journal} {J. Chem. Phys.}\ }\textbf {\bibinfo {volume} {155}},\ \bibinfo
  {pages} {210901} (\bibinfo {year} {2021})}\BibitemShut {NoStop}%
\bibitem [{\citenamefont {Pollak}, \citenamefont {Grabert},\ and\ \citenamefont
  {H{\"a}nggi}(1989)}]{pgh89}%
  \BibitemOpen
  \bibfield  {author} {\bibinfo {author} {\bibfnamefont {E.}~\bibnamefont
  {Pollak}}, \bibinfo {author} {\bibfnamefont {H.}~\bibnamefont {Grabert}},\
  and\ \bibinfo {author} {\bibfnamefont {P.}~\bibnamefont {H{\"a}nggi}},\
  }\bibfield  {title} {\enquote {\bibinfo {title} {Theory of activated rate
  processes for arbitrary frequency dependent friction: Solution of the
  turnover problem},}\ }\href {https://doi.org/10.1063/1.456837} {\bibfield
  {journal} {\bibinfo  {journal} {J. Chem. Phys.}\ }\textbf {\bibinfo {volume}
  {91}},\ \bibinfo {pages} {4073--4087} (\bibinfo {year} {1989})}\BibitemShut
  {NoStop}%
\bibitem [{\citenamefont {Kramers}(1940)}]{kram40}%
  \BibitemOpen
  \bibfield  {author} {\bibinfo {author} {\bibfnamefont {H.~A.}\ \bibnamefont
  {Kramers}},\ }\bibfield  {title} {\enquote {\bibinfo {title} {Brownian motion
  in a field of force and the diffusional model of chemical reactions},}\
  }\href {https://doi.org/10.1016/S0031-8914(40)90098-2} {\bibfield  {journal}
  {\bibinfo  {journal} {Physica (Utrecht)}\ }\textbf {\bibinfo {volume} {7}},\
  \bibinfo {pages} {284--304} (\bibinfo {year} {1940})}\BibitemShut {NoStop}%
\bibitem [{\citenamefont {{Garc{\'\i}a-M\"uller}}\ \emph
  {et~al.}(2008)\citenamefont {{Garc{\'\i}a-M\"uller}}, \citenamefont
  {Borondo}, \citenamefont {Hernandez},\ and\ \citenamefont
  {Benito}}]{hern08g}%
  \BibitemOpen
  \bibfield  {author} {\bibinfo {author} {\bibfnamefont {P.~L.}\ \bibnamefont
  {{Garc{\'\i}a-M\"uller}}}, \bibinfo {author} {\bibfnamefont {F.}~\bibnamefont
  {Borondo}}, \bibinfo {author} {\bibfnamefont {R.}~\bibnamefont {Hernandez}},\
  and\ \bibinfo {author} {\bibfnamefont {R.~M.}\ \bibnamefont {Benito}},\
  }\bibfield  {title} {\enquote {\bibinfo {title} {Solvent-induced acceleration
  of the rate of activation of a molecular reaction},}\ }\href
  {https://doi.org/10.1103/PhysRevLett.101.178302} {\bibfield  {journal}
  {\bibinfo  {journal} {Phys. Rev. Lett.}\ }\textbf {\bibinfo {volume} {101}},\
  \bibinfo {pages} {178302--01--04} (\bibinfo {year} {2008})}\BibitemShut
  {NoStop}%
\bibitem [{\citenamefont {Pellouchoud}\ and\ \citenamefont
  {Reed}(2015)}]{Pellouchoud2015a}%
  \BibitemOpen
  \bibfield  {author} {\bibinfo {author} {\bibfnamefont {L.~A.}\ \bibnamefont
  {Pellouchoud}}\ and\ \bibinfo {author} {\bibfnamefont {E.~J.}\ \bibnamefont
  {Reed}},\ }\bibfield  {title} {\enquote {\bibinfo {title} {Coherent chemistry
  with {THz} pulses: Ultrafast field-driven isomerization of {LiNC}},}\ }\href
  {https://doi.org/10.1103/PhysRevA.91.052706} {\bibfield  {journal} {\bibinfo
  {journal} {Phys. Rev. A}\ }\textbf {\bibinfo {volume} {91}},\ \bibinfo
  {pages} {052706} (\bibinfo {year} {2015})}\BibitemShut {NoStop}%
\bibitem [{\citenamefont {Kubo}(1966)}]{kubo66}%
  \BibitemOpen
  \bibfield  {author} {\bibinfo {author} {\bibfnamefont {R.}~\bibnamefont
  {Kubo}},\ }\bibfield  {title} {\enquote {\bibinfo {title} {The
  fluctuation-dissipation theorem},}\ }\href
  {https://doi.org/10.1088/0034-4885/29/1/306} {\bibfield  {journal} {\bibinfo
  {journal} {Rep. Prog. Phys.}\ }\textbf {\bibinfo {volume} {29}},\ \bibinfo
  {pages} {255--284} (\bibinfo {year} {1966})}\BibitemShut {NoStop}%
\bibitem [{\citenamefont {Hanggi}(1986)}]{Hanggi1986a}%
  \BibitemOpen
  \bibfield  {author} {\bibinfo {author} {\bibfnamefont {P.}~\bibnamefont
  {Hanggi}},\ }\bibfield  {title} {\enquote {\bibinfo {title} {Escape from a
  metastable state},}\ }\href {https://doi.org/10.1007/bf01010843} {\bibfield
  {journal} {\bibinfo  {journal} {J. Stat. Phys.}\ }\textbf {\bibinfo {volume}
  {42}},\ \bibinfo {pages} {105--148} (\bibinfo {year} {1986})}\BibitemShut
  {NoStop}%
\bibitem [{\citenamefont {Straub}\ and\ \citenamefont
  {Berne}(1986)}]{Straub1986a}%
  \BibitemOpen
  \bibfield  {author} {\bibinfo {author} {\bibfnamefont {J.~E.}\ \bibnamefont
  {Straub}}\ and\ \bibinfo {author} {\bibfnamefont {B.~J.}\ \bibnamefont
  {Berne}},\ }\bibfield  {title} {\enquote {\bibinfo {title} {Energy diffusion
  in many-dimensional {Markovian} systems: The consequences of competition
  between inter- and intramolecular vibrational energy transfer},}\ }\href
  {https://doi.org/10.1063/1.451009} {\bibfield  {journal} {\bibinfo  {journal}
  {J. Chem. Phys.}\ }\textbf {\bibinfo {volume} {85}},\ \bibinfo {pages}
  {2999--3006} (\bibinfo {year} {1986})}\BibitemShut {NoStop}%
\bibitem [{\citenamefont {Zwanzig}(1987)}]{Zwanzig1987a}%
  \BibitemOpen
  \bibfield  {author} {\bibinfo {author} {\bibfnamefont {R.}~\bibnamefont
  {Zwanzig}},\ }\bibfield  {title} {\enquote {\bibinfo {title} {Comments on a
  paper by {Straub}, {Borkovec}, and {Berne}},}\ }\href
  {https://doi.org/10.1063/1.452509} {\bibfield  {journal} {\bibinfo  {journal}
  {J. Chem. Phys.}\ }\textbf {\bibinfo {volume} {86}},\ \bibinfo {pages}
  {5801--5803} (\bibinfo {year} {1987})}\BibitemShut {NoStop}%
\bibitem [{\citenamefont {Berne}, \citenamefont {Borkovec},\ and\ \citenamefont
  {Straub}(1988)}]{berne88}%
  \BibitemOpen
  \bibfield  {author} {\bibinfo {author} {\bibfnamefont {B.~J.}\ \bibnamefont
  {Berne}}, \bibinfo {author} {\bibfnamefont {M.}~\bibnamefont {Borkovec}},\
  and\ \bibinfo {author} {\bibfnamefont {J.~E.}\ \bibnamefont {Straub}},\
  }\bibfield  {title} {\enquote {\bibinfo {title} {Classical and modern methods
  in reaction rate theory},}\ }\href {https://doi.org/10.1021/j100324a007}
  {\bibfield  {journal} {\bibinfo  {journal} {J. Phys. Chem.}\ }\textbf
  {\bibinfo {volume} {92}},\ \bibinfo {pages} {3711--3725} (\bibinfo {year}
  {1988})}\BibitemShut {NoStop}%
\bibitem [{\citenamefont {H{\"a}nggi}, \citenamefont {Talkner},\ and\
  \citenamefont {Borkovec}(1990)}]{rmp90}%
  \BibitemOpen
  \bibfield  {author} {\bibinfo {author} {\bibfnamefont {P.}~\bibnamefont
  {H{\"a}nggi}}, \bibinfo {author} {\bibfnamefont {P.}~\bibnamefont
  {Talkner}},\ and\ \bibinfo {author} {\bibfnamefont {M.}~\bibnamefont
  {Borkovec}},\ }\bibfield  {title} {\enquote {\bibinfo {title} {Reaction-rate
  theory: Fifty years after {Kramers}},}\ }\href
  {https://doi.org/10.1103/RevModPhys.62.251} {\bibfield  {journal} {\bibinfo
  {journal} {Rev. Mod. Phys.}\ }\textbf {\bibinfo {volume} {62}},\ \bibinfo
  {pages} {251--341} (\bibinfo {year} {1990})},\ \bibinfo {note} {and
  references therein}\BibitemShut {NoStop}%
\bibitem [{\citenamefont {Straub}, \citenamefont {Borkovec},\ and\
  \citenamefont {Berne}(1985)}]{sbb85}%
  \BibitemOpen
  \bibfield  {author} {\bibinfo {author} {\bibfnamefont {J.~E.}\ \bibnamefont
  {Straub}}, \bibinfo {author} {\bibfnamefont {M.}~\bibnamefont {Borkovec}},\
  and\ \bibinfo {author} {\bibfnamefont {B.~J.}\ \bibnamefont {Berne}},\
  }\bibfield  {title} {\enquote {\bibinfo {title} {Shortcomings of current
  theories of non-{Markovian} activated rate processes},}\ }\href
  {https://doi.org/10.1063/1.449172} {\bibfield  {journal} {\bibinfo  {journal}
  {J. Chem. Phys.}\ }\textbf {\bibinfo {volume} {83}},\ \bibinfo {pages}
  {3172--4} (\bibinfo {year} {1985})}\BibitemShut {NoStop}%
\bibitem [{\citenamefont {Straub}, \citenamefont {Borkovec},\ and\
  \citenamefont {Berne}(1986)}]{sbb86}%
  \BibitemOpen
  \bibfield  {author} {\bibinfo {author} {\bibfnamefont {J.~E.}\ \bibnamefont
  {Straub}}, \bibinfo {author} {\bibfnamefont {M.}~\bibnamefont {Borkovec}},\
  and\ \bibinfo {author} {\bibfnamefont {B.~J.}\ \bibnamefont {Berne}},\
  }\bibfield  {title} {\enquote {\bibinfo {title} {Non-{Markovian} activated
  rate processes: Comparison of current theories with numerical simulation
  data},}\ }\href {https://doi.org/10.1063/1.450425} {\bibfield  {journal}
  {\bibinfo  {journal} {J. Chem. Phys.}\ }\textbf {\bibinfo {volume} {84}},\
  \bibinfo {pages} {1788--1794} (\bibinfo {year} {1986})}\BibitemShut {NoStop}%
\bibitem [{\citenamefont {Hill}(1989)}]{Hill1989a}%
  \BibitemOpen
  \bibfield  {author} {\bibinfo {author} {\bibfnamefont {T.~L.}\ \bibnamefont
  {Hill}},\ }\href {https://doi.org/10.1007/978-1-4612-3558-3} {\emph {\bibinfo
  {title} {Free Energy Transduction and Biochemical Cycle Kinetics}}}\
  (\bibinfo  {publisher} {Springer New York},\ \bibinfo {year}
  {1989})\BibitemShut {NoStop}%
\bibitem [{\citenamefont {Eyring}(1935{\natexlab{a}})}]{eyring35}%
  \BibitemOpen
  \bibfield  {author} {\bibinfo {author} {\bibfnamefont {H.}~\bibnamefont
  {Eyring}},\ }\bibfield  {title} {\enquote {\bibinfo {title} {The activated
  complex in chemical reactions},}\ }\href {https://doi.org/10.1063/1.1749604}
  {\bibfield  {journal} {\bibinfo  {journal} {J. Chem. Phys.}\ }\textbf
  {\bibinfo {volume} {3}},\ \bibinfo {pages} {107--115} (\bibinfo {year}
  {1935}{\natexlab{a}})}\BibitemShut {NoStop}%
\bibitem [{\citenamefont {Wigner}(1937)}]{wigner37}%
  \BibitemOpen
  \bibfield  {author} {\bibinfo {author} {\bibfnamefont {E.~P.}\ \bibnamefont
  {Wigner}},\ }\bibfield  {title} {\enquote {\bibinfo {title} {Calculation of
  the rate of elementary association reactions},}\ }\href
  {https://doi.org/10.1063/1.1750107} {\bibfield  {journal} {\bibinfo
  {journal} {J. Chem. Phys.}\ }\textbf {\bibinfo {volume} {5}},\ \bibinfo
  {pages} {720--725} (\bibinfo {year} {1937})}\BibitemShut {NoStop}%
\bibitem [{\citenamefont {Pechukas}(1981)}]{pech81}%
  \BibitemOpen
  \bibfield  {author} {\bibinfo {author} {\bibfnamefont {P.}~\bibnamefont
  {Pechukas}},\ }\bibfield  {title} {\enquote {\bibinfo {title} {Transition
  state theory},}\ }\href {https://doi.org/10.1146/annurev.pc.32.100181.001111}
  {\bibfield  {journal} {\bibinfo  {journal} {Annu. Rev. Phys. Chem.}\ }\textbf
  {\bibinfo {volume} {32}},\ \bibinfo {pages} {159--177} (\bibinfo {year}
  {1981})}\BibitemShut {NoStop}%
\bibitem [{\citenamefont {Truhlar}, \citenamefont {Garrett},\ and\
  \citenamefont {Klippenstein}(1996)}]{truh96}%
  \BibitemOpen
  \bibfield  {author} {\bibinfo {author} {\bibfnamefont {D.~G.}\ \bibnamefont
  {Truhlar}}, \bibinfo {author} {\bibfnamefont {B.~C.}\ \bibnamefont
  {Garrett}},\ and\ \bibinfo {author} {\bibfnamefont {S.~J.}\ \bibnamefont
  {Klippenstein}},\ }\bibfield  {title} {\enquote {\bibinfo {title} {Current
  status of transition-state theory},}\ }\href
  {https://doi.org/10.1021/jp953748q} {\bibfield  {journal} {\bibinfo
  {journal} {J. Phys. Chem.}\ }\textbf {\bibinfo {volume} {100}},\ \bibinfo
  {pages} {12771--12800} (\bibinfo {year} {1996})}\BibitemShut {NoStop}%
\bibitem [{\citenamefont {Hernandez}, \citenamefont {Bartsch},\ and\
  \citenamefont {Uzer}(2010)}]{hern10a}%
  \BibitemOpen
  \bibfield  {author} {\bibinfo {author} {\bibfnamefont {R.}~\bibnamefont
  {Hernandez}}, \bibinfo {author} {\bibfnamefont {T.}~\bibnamefont {Bartsch}},\
  and\ \bibinfo {author} {\bibfnamefont {T.}~\bibnamefont {Uzer}},\ }\bibfield
  {title} {\enquote {\bibinfo {title} {Transition state theory in liquids
  beyond planar dividing surfaces},}\ }\href
  {https://doi.org/10.1016/j.chemphys.2010.01.016} {\bibfield  {journal}
  {\bibinfo  {journal} {Chem. Phys.}\ }\textbf {\bibinfo {volume} {370}},\
  \bibinfo {pages} {270--276} (\bibinfo {year} {2010})}\BibitemShut {NoStop}%
\bibitem [{\citenamefont {Mullen}, \citenamefont {Shea},\ and\ \citenamefont
  {Peters}(2014)}]{peters14a}%
  \BibitemOpen
  \bibfield  {author} {\bibinfo {author} {\bibfnamefont {R.~G.}\ \bibnamefont
  {Mullen}}, \bibinfo {author} {\bibfnamefont {J.-E.}\ \bibnamefont {Shea}},\
  and\ \bibinfo {author} {\bibfnamefont {B.}~\bibnamefont {Peters}},\
  }\bibfield  {title} {\enquote {\bibinfo {title} {Communication: An existence
  test for dividing surfaces without recrossing},}\ }\href
  {https://doi.org/10.1063/1.4862504} {\bibfield  {journal} {\bibinfo
  {journal} {J. Chem. Phys.}\ }\textbf {\bibinfo {volume} {140}},\ \bibinfo
  {pages} {041104} (\bibinfo {year} {2014})}\BibitemShut {NoStop}%
\bibitem [{\citenamefont {Wiggins}(2016)}]{wiggins16}%
  \BibitemOpen
  \bibfield  {author} {\bibinfo {author} {\bibfnamefont {S.}~\bibnamefont
  {Wiggins}},\ }\bibfield  {title} {\enquote {\bibinfo {title} {The role of
  normally hyperbolic invariant manifolds ({NHIMS}) in the context of the phase
  space setting for chemical reaction dynamics},}\ }\href
  {https://doi.org/10.1134/S1560354716060034} {\bibfield  {journal} {\bibinfo
  {journal} {Regul. Chaotic Dyn.}\ }\textbf {\bibinfo {volume} {21}},\ \bibinfo
  {pages} {621--638} (\bibinfo {year} {2016})}\BibitemShut {NoStop}%
\bibitem [{\citenamefont {Ezra}\ and\ \citenamefont
  {Wiggins}(2018)}]{Ezra2018a}%
  \BibitemOpen
  \bibfield  {author} {\bibinfo {author} {\bibfnamefont {G.~S.}\ \bibnamefont
  {Ezra}}\ and\ \bibinfo {author} {\bibfnamefont {S.}~\bibnamefont {Wiggins}},\
  }\bibfield  {title} {\enquote {\bibinfo {title} {Sampling phase space
  dividing surfaces constructed from normally hyperbolic invariant manifolds
  ({NHIMs})},}\ }\href {https://doi.org/10.1021/acs.jpca.8b07205} {\bibfield
  {journal} {\bibinfo  {journal} {J. Phys. Chem. A}\ }\textbf {\bibinfo
  {volume} {122}},\ \bibinfo {pages} {8354--8362} (\bibinfo {year}
  {2018})}\BibitemShut {NoStop}%
\bibitem [{\citenamefont {Talkner}(1987)}]{talkner87a}%
  \BibitemOpen
  \bibfield  {author} {\bibinfo {author} {\bibfnamefont {P.}~\bibnamefont
  {Talkner}},\ }\bibfield  {title} {\enquote {\bibinfo {title} {Mean first
  passage time and the lifetime of a metastable state},}\ }\href@noop {}
  {\bibfield  {journal} {\bibinfo  {journal} {Z. Physik B}\ }\textbf {\bibinfo
  {volume} {68}},\ \bibinfo {pages} {201--207} (\bibinfo {year}
  {1987})}\BibitemShut {NoStop}%
\bibitem [{\citenamefont {Lee}\ and\ \citenamefont
  {Karplus}(1988)}]{karplus88}%
  \BibitemOpen
  \bibfield  {author} {\bibinfo {author} {\bibfnamefont {S.}~\bibnamefont
  {Lee}}\ and\ \bibinfo {author} {\bibfnamefont {M.}~\bibnamefont {Karplus}},\
  }\bibfield  {title} {\enquote {\bibinfo {title} {Dynamics of reactions
  involving diffusive multidimensional barrier crossing},}\ }\href
  {https://doi.org/10.1021/j100316a018} {\bibfield  {journal} {\bibinfo
  {journal} {J. Phys. Chem.}\ }\textbf {\bibinfo {volume} {92}},\ \bibinfo
  {pages} {1075--1086} (\bibinfo {year} {1988})}\BibitemShut {NoStop}%
\bibitem [{\citenamefont {Mel'nikov}\ and\ \citenamefont
  {Meshkov}(1986)}]{mm86}%
  \BibitemOpen
  \bibfield  {author} {\bibinfo {author} {\bibfnamefont {V.~I.}\ \bibnamefont
  {Mel'nikov}}\ and\ \bibinfo {author} {\bibfnamefont {S.~V.}\ \bibnamefont
  {Meshkov}},\ }\bibfield  {title} {\enquote {\bibinfo {title} {Theory of
  activated rate processes: Exact solution of the {K}ramers problem},}\ }\href
  {https://doi.org/10.1063/1.451844} {\bibfield  {journal} {\bibinfo  {journal}
  {J. Chem. Phys.}\ }\textbf {\bibinfo {volume} {85}},\ \bibinfo {pages}
  {1018--1027} (\bibinfo {year} {1986})}\BibitemShut {NoStop}%
\bibitem [{\citenamefont {Ianconescu}\ and\ \citenamefont
  {Pollak}(2016)}]{pollak16}%
  \BibitemOpen
  \bibfield  {author} {\bibinfo {author} {\bibfnamefont {R.}~\bibnamefont
  {Ianconescu}}\ and\ \bibinfo {author} {\bibfnamefont {E.}~\bibnamefont
  {Pollak}},\ }\bibfield  {title} {\enquote {\bibinfo {title} {Kramers’
  turnover theory: Improvement and extension to low barriers},}\ }\href
  {https://doi.org/10.1021/acs.jpca.5b11502} {\bibfield  {journal} {\bibinfo
  {journal} {J. Phys. Chem. A}\ }\textbf {\bibinfo {volume} {120}},\ \bibinfo
  {pages} {3155--3164} (\bibinfo {year} {2016})}\BibitemShut {NoStop}%
\bibitem [{\citenamefont {Levine}\ and\ \citenamefont
  {Bernstein}(1987)}]{bernstein87}%
  \BibitemOpen
  \bibfield  {author} {\bibinfo {author} {\bibfnamefont {R.~D.}\ \bibnamefont
  {Levine}}\ and\ \bibinfo {author} {\bibfnamefont {R.~B.}\ \bibnamefont
  {Bernstein}},\ }\href@noop {} {\emph {\bibinfo {title} {Molecular Reaction
  Dynamics and Chemical Reactivity}}}\ (\bibinfo  {publisher} {Oxford
  University Press},\ \bibinfo {address} {New York},\ \bibinfo {year}
  {1987})\BibitemShut {NoStop}%
\bibitem [{\citenamefont {Steinfeld}, \citenamefont {Francisco},\ and\
  \citenamefont {Hase}(1999)}]{hasebook99}%
  \BibitemOpen
  \bibfield  {author} {\bibinfo {author} {\bibfnamefont {J.~I.}\ \bibnamefont
  {Steinfeld}}, \bibinfo {author} {\bibfnamefont {J.~S.}\ \bibnamefont
  {Francisco}},\ and\ \bibinfo {author} {\bibfnamefont {W.~L.}\ \bibnamefont
  {Hase}},\ }\href@noop {} {\emph {\bibinfo {title} {Chemical Kinetics and
  Dynamics}}},\ \bibinfo {edition} {2nd}\ ed.\ (\bibinfo  {publisher} {Prentice
  Hall},\ \bibinfo {address} {Upper Saddle River, NJ},\ \bibinfo {year}
  {1999})\BibitemShut {NoStop}%
\bibitem [{\citenamefont {Uhlenbeck}\ and\ \citenamefont
  {Ornstein}(1930)}]{Uhlenbeck1930a}%
  \BibitemOpen
  \bibfield  {author} {\bibinfo {author} {\bibfnamefont {G.~E.}\ \bibnamefont
  {Uhlenbeck}}\ and\ \bibinfo {author} {\bibfnamefont {L.~S.}\ \bibnamefont
  {Ornstein}},\ }\bibfield  {title} {\enquote {\bibinfo {title} {On the theory
  of the {Brownian} motion},}\ }\href {https://doi.org/10.1103/PhysRev.36.823}
  {\bibfield  {journal} {\bibinfo  {journal} {Phys. Rev.}\ }\textbf {\bibinfo
  {volume} {36}},\ \bibinfo {pages} {823--841} (\bibinfo {year}
  {1930})}\BibitemShut {NoStop}%
\bibitem [{\citenamefont {Kappler}\ \emph {et~al.}(2018)\citenamefont
  {Kappler}, \citenamefont {Daldrop}, \citenamefont {Brünig}, \citenamefont
  {Boehle},\ and\ \citenamefont {Netz}}]{Kappler2018a}%
  \BibitemOpen
  \bibfield  {author} {\bibinfo {author} {\bibfnamefont {J.}~\bibnamefont
  {Kappler}}, \bibinfo {author} {\bibfnamefont {J.~O.}\ \bibnamefont
  {Daldrop}}, \bibinfo {author} {\bibfnamefont {F.~N.}\ \bibnamefont
  {Brünig}}, \bibinfo {author} {\bibfnamefont {M.~D.}\ \bibnamefont
  {Boehle}},\ and\ \bibinfo {author} {\bibfnamefont {R.~R.}\ \bibnamefont
  {Netz}},\ }\bibfield  {title} {\enquote {\bibinfo {title} {Memory-induced
  acceleration and slowdown of barrier crossing},}\ }\href
  {https://doi.org/10.1063/1.4998239} {\bibfield  {journal} {\bibinfo
  {journal} {J. Chem. Phys.}\ }\textbf {\bibinfo {volume} {148}},\ \bibinfo
  {pages} {014903} (\bibinfo {year} {2018})}\BibitemShut {NoStop}%
\bibitem [{\citenamefont {Shepherd}\ and\ \citenamefont
  {Hernandez}(2001)}]{hern01d}%
  \BibitemOpen
  \bibfield  {author} {\bibinfo {author} {\bibfnamefont {T.~D.}\ \bibnamefont
  {Shepherd}}\ and\ \bibinfo {author} {\bibfnamefont {R.}~\bibnamefont
  {Hernandez}},\ }\bibfield  {title} {\enquote {\bibinfo {title} {Chemical
  reaction dynamics with stochastic potentials beyond the high-friction
  limit},}\ }\href {https://doi.org/10.1063/1.1386422} {\bibfield  {journal}
  {\bibinfo  {journal} {J. Chem. Phys.}\ }\textbf {\bibinfo {volume} {115}},\
  \bibinfo {pages} {2430--2438} (\bibinfo {year} {2001})}\BibitemShut {NoStop}%
\bibitem [{\citenamefont {Polanyi}\ and\ \citenamefont
  {Zewail}(1995)}]{Polanyi1995a}%
  \BibitemOpen
  \bibfield  {author} {\bibinfo {author} {\bibfnamefont {J.~C.}\ \bibnamefont
  {Polanyi}}\ and\ \bibinfo {author} {\bibfnamefont {A.~H.}\ \bibnamefont
  {Zewail}},\ }\bibfield  {title} {\enquote {\bibinfo {title} {Direct
  observation of the transition state},}\ }\href
  {https://doi.org/10.1021/ar00051a005} {\bibfield  {journal} {\bibinfo
  {journal} {Acc. Chem. Res.}\ }\textbf {\bibinfo {volume} {28}},\ \bibinfo
  {pages} {119--132} (\bibinfo {year} {1995})}\BibitemShut {NoStop}%
\bibitem [{\citenamefont {Wenthold}\ \emph {et~al.}(1996)\citenamefont
  {Wenthold}, \citenamefont {Hrovat}, \citenamefont {Borden},\ and\
  \citenamefont {Lineberger}}]{Wenthold1996a}%
  \BibitemOpen
  \bibfield  {author} {\bibinfo {author} {\bibfnamefont {P.~G.}\ \bibnamefont
  {Wenthold}}, \bibinfo {author} {\bibfnamefont {D.~A.}\ \bibnamefont
  {Hrovat}}, \bibinfo {author} {\bibfnamefont {W.~T.}\ \bibnamefont {Borden}},\
  and\ \bibinfo {author} {\bibfnamefont {W.~C.}\ \bibnamefont {Lineberger}},\
  }\bibfield  {title} {\enquote {\bibinfo {title} {Transition-state
  spectroscopy of cyclooctatetraene},}\ }\href
  {https://doi.org/10.1126/science.272.5267.1456} {\bibfield  {journal}
  {\bibinfo  {journal} {Science}\ }\textbf {\bibinfo {volume} {272}},\ \bibinfo
  {pages} {1456--1459} (\bibinfo {year} {1996})}\BibitemShut {NoStop}%
\bibitem [{\citenamefont {Neumark}(1996)}]{Neumark1996a}%
  \BibitemOpen
  \bibfield  {author} {\bibinfo {author} {\bibfnamefont {D.~M.}\ \bibnamefont
  {Neumark}},\ }\bibfield  {title} {\enquote {\bibinfo {title} {Transition
  state spectroscopy},}\ }\href {https://doi.org/10.1126/science.272.5267.1446}
  {\bibfield  {journal} {\bibinfo  {journal} {Science}\ }\textbf {\bibinfo
  {volume} {272}},\ \bibinfo {pages} {1446--1447} (\bibinfo {year}
  {1996})}\BibitemShut {NoStop}%
\bibitem [{\citenamefont {Hamm}\ and\ \citenamefont {Zanni}(2009)}]{Hamm2009a}%
  \BibitemOpen
  \bibfield  {author} {\bibinfo {author} {\bibfnamefont {P.}~\bibnamefont
  {Hamm}}\ and\ \bibinfo {author} {\bibfnamefont {M.}~\bibnamefont {Zanni}},\
  }\href {https://doi.org/10.1017/cbo9780511675935} {\emph {\bibinfo {title}
  {Concepts and Methods of {2D} Infrared Spectroscopy}}}\ (\bibinfo
  {publisher} {Cambridge University Press},\ \bibinfo {year}
  {2009})\BibitemShut {NoStop}%
\bibitem [{\citenamefont {Truhlar}, \citenamefont {Hase},\ and\ \citenamefont
  {Hynes}(1983)}]{truh83}%
  \BibitemOpen
  \bibfield  {author} {\bibinfo {author} {\bibfnamefont {D.~G.}\ \bibnamefont
  {Truhlar}}, \bibinfo {author} {\bibfnamefont {W.~L.}\ \bibnamefont {Hase}},\
  and\ \bibinfo {author} {\bibfnamefont {J.~T.}\ \bibnamefont {Hynes}},\
  }\bibfield  {title} {\enquote {\bibinfo {title} {Current status of
  transition--state theory},}\ }\href {https://doi.org/10.1021/j100238a003}
  {\bibfield  {journal} {\bibinfo  {journal} {J. Phys. Chem.}\ }\textbf
  {\bibinfo {volume} {87}},\ \bibinfo {pages} {2664--2682} (\bibinfo {year}
  {1983})}\BibitemShut {NoStop}%
\bibitem [{\citenamefont {Hynes}(1985)}]{hynes85b}%
  \BibitemOpen
  \bibfield  {author} {\bibinfo {author} {\bibfnamefont {J.~T.}\ \bibnamefont
  {Hynes}},\ }\bibfield  {title} {\enquote {\bibinfo {title} {Chemical reaction
  dynamics in solution},}\ }\href
  {https://doi.org/10.1146/annurev.pc.36.100185.003041} {\bibfield  {journal}
  {\bibinfo  {journal} {Annu. Rev. Phys. Chem.}\ }\textbf {\bibinfo {volume}
  {36}},\ \bibinfo {pages} {573--597} (\bibinfo {year} {1985})}\BibitemShut
  {NoStop}%
\bibitem [{\citenamefont {Straub}, \citenamefont {Borkovec},\ and\
  \citenamefont {Berne}(1988)}]{sbb88}%
  \BibitemOpen
  \bibfield  {author} {\bibinfo {author} {\bibfnamefont {J.~E.}\ \bibnamefont
  {Straub}}, \bibinfo {author} {\bibfnamefont {M.}~\bibnamefont {Borkovec}},\
  and\ \bibinfo {author} {\bibfnamefont {B.~J.}\ \bibnamefont {Berne}},\
  }\bibfield  {title} {\enquote {\bibinfo {title} {Molecular dynamics study of
  an isomerizing diatomic in a lennard-jones fluid},}\ }\href
  {https://doi.org/10.1063/1.455678} {\bibfield  {journal} {\bibinfo  {journal}
  {J. Chem. Phys.}\ }\textbf {\bibinfo {volume} {89}},\ \bibinfo {pages}
  {4833--4847} (\bibinfo {year} {1988})}\BibitemShut {NoStop}%
\bibitem [{\citenamefont {Tucker}\ \emph {et~al.}(1991)\citenamefont {Tucker},
  \citenamefont {Tuckerman}, \citenamefont {Berne},\ and\ \citenamefont
  {Pollak}}]{tucker91}%
  \BibitemOpen
  \bibfield  {author} {\bibinfo {author} {\bibfnamefont {S.~C.}\ \bibnamefont
  {Tucker}}, \bibinfo {author} {\bibfnamefont {M.~E.}\ \bibnamefont
  {Tuckerman}}, \bibinfo {author} {\bibfnamefont {B.~J.}\ \bibnamefont
  {Berne}},\ and\ \bibinfo {author} {\bibfnamefont {E.}~\bibnamefont
  {Pollak}},\ }\bibfield  {title} {\enquote {\bibinfo {title} {Comparison of
  rate theories for generalized {L}angevin dynamics},}\ }\href
  {https://doi.org/10.1063/1.461603} {\bibfield  {journal} {\bibinfo  {journal}
  {J. Chem. Phys.}\ }\textbf {\bibinfo {volume} {95}},\ \bibinfo {pages} {5809}
  (\bibinfo {year} {1991})}\BibitemShut {NoStop}%
\bibitem [{\citenamefont {Bolhuis}, \citenamefont {Dellago},\ and\
  \citenamefont {Chandler}(2000)}]{chandler00}%
  \BibitemOpen
  \bibfield  {author} {\bibinfo {author} {\bibfnamefont {P.~G.}\ \bibnamefont
  {Bolhuis}}, \bibinfo {author} {\bibfnamefont {C.~P.}\ \bibnamefont
  {Dellago}},\ and\ \bibinfo {author} {\bibfnamefont {D.}~\bibnamefont
  {Chandler}},\ }\bibfield  {title} {\enquote {\bibinfo {title} {Reaction
  coordinates of biomolecular isomerization},}\ }\href@noop {} {\bibfield
  {journal} {\bibinfo  {journal} {Proc. Natl. Acad. Sci. U.S.A.}\ }\textbf
  {\bibinfo {volume} {97}},\ \bibinfo {pages} {5877--5882} (\bibinfo {year}
  {2000})}\BibitemShut {NoStop}%
\bibitem [{\citenamefont {Truhlar}\ and\ \citenamefont
  {Garrett}(2000)}]{truh2000}%
  \BibitemOpen
  \bibfield  {author} {\bibinfo {author} {\bibfnamefont {D.~G.}\ \bibnamefont
  {Truhlar}}\ and\ \bibinfo {author} {\bibfnamefont {B.~C.}\ \bibnamefont
  {Garrett}},\ }\bibfield  {title} {\enquote {\bibinfo {title}
  {Multidimensional transition state theory and the validity of {Grote-Hynes}
  theory},}\ }\href {https://doi.org/10.1021/jp992430l} {\bibfield  {journal}
  {\bibinfo  {journal} {J. Phys. Chem. B}\ }\textbf {\bibinfo {volume} {104}},\
  \bibinfo {pages} {1069--1072} (\bibinfo {year} {2000})}\BibitemShut {NoStop}%
\bibitem [{\citenamefont {Tiwary}\ and\ \citenamefont
  {Berne}(2016)}]{Berne2016}%
  \BibitemOpen
  \bibfield  {author} {\bibinfo {author} {\bibfnamefont {P.}~\bibnamefont
  {Tiwary}}\ and\ \bibinfo {author} {\bibfnamefont {B.~J.}\ \bibnamefont
  {Berne}},\ }\bibfield  {title} {\enquote {\bibinfo {title} {Kramers turnover:
  From energy diffusion to spatial diffusion using metadynamics},}\ }\href
  {https://doi.org/10.1063/1.4944577} {\bibfield  {journal} {\bibinfo
  {journal} {J. Chem. Phys.}\ }\textbf {\bibinfo {volume} {144}},\ \bibinfo
  {pages} {134103} (\bibinfo {year} {2016})}\BibitemShut {NoStop}%
\bibitem [{\citenamefont {Grote}\ and\ \citenamefont
  {Hynes}(1980)}]{Grote-1980}%
  \BibitemOpen
  \bibfield  {author} {\bibinfo {author} {\bibfnamefont {R.~F.}\ \bibnamefont
  {Grote}}\ and\ \bibinfo {author} {\bibfnamefont {J.~T.}\ \bibnamefont
  {Hynes}},\ }\href@noop {} {\bibfield  {journal} {\bibinfo  {journal} {J.
  Chem. Phys.}\ }\textbf {\bibinfo {volume} {73}},\ \bibinfo {pages} {2715}
  (\bibinfo {year} {1980})}\BibitemShut {NoStop}%
\bibitem [{\citenamefont {Pollak}(1986)}]{pollak86b}%
  \BibitemOpen
  \bibfield  {author} {\bibinfo {author} {\bibfnamefont {E.}~\bibnamefont
  {Pollak}},\ }\bibfield  {title} {\enquote {\bibinfo {title} {Theory of
  activated rate processes: A new derivation of {K}ramers' expression},}\
  }\href {https://doi.org/10.1063/1.451294} {\bibfield  {journal} {\bibinfo
  {journal} {J. Chem. Phys.}\ }\textbf {\bibinfo {volume} {85}},\ \bibinfo
  {pages} {865--867} (\bibinfo {year} {1986})}\BibitemShut {NoStop}%
\bibitem [{\citenamefont {Peters}(2017)}]{Peters2017a}%
  \BibitemOpen
  \bibfield  {author} {\bibinfo {author} {\bibfnamefont {B.}~\bibnamefont
  {Peters}},\ }\href@noop {} {\emph {\bibinfo {title} {Reaction Rate Theory and
  Rare Events Simulations}}}\ (\bibinfo  {publisher} {Elsevier},\ \bibinfo
  {address} {Amsterdam},\ \bibinfo {year} {2017})\BibitemShut {NoStop}%
\bibitem [{\citenamefont {Peters}, \citenamefont {Bell},\ and\ \citenamefont
  {Chakraborty}(2004)}]{Peters04a}%
  \BibitemOpen
  \bibfield  {author} {\bibinfo {author} {\bibfnamefont {B.}~\bibnamefont
  {Peters}}, \bibinfo {author} {\bibfnamefont {A.~T.}\ \bibnamefont {Bell}},\
  and\ \bibinfo {author} {\bibfnamefont {A.}~\bibnamefont {Chakraborty}},\
  }\bibfield  {title} {\enquote {\bibinfo {title} {Rate constants from the
  reaction path hamiltonian. i. reactive flux simulations for dynamically
  correct rates},}\ }\href {https://doi.org/10.1063/1.1778161} {\bibfield
  {journal} {\bibinfo  {journal} {J. Chem. Phys.}\ }\textbf {\bibinfo {volume}
  {121}},\ \bibinfo {pages} {4453--4460} (\bibinfo {year} {2004})}\BibitemShut
  {NoStop}%
\bibitem [{\citenamefont {Ciccotti}\ \emph {et~al.}(1990)\citenamefont
  {Ciccotti}, \citenamefont {Ferrario}, \citenamefont {Hynes},\ and\
  \citenamefont {Kapral}}]{Ciccotti1990a}%
  \BibitemOpen
  \bibfield  {author} {\bibinfo {author} {\bibfnamefont {G.}~\bibnamefont
  {Ciccotti}}, \bibinfo {author} {\bibfnamefont {M.}~\bibnamefont {Ferrario}},
  \bibinfo {author} {\bibfnamefont {J.~T.}\ \bibnamefont {Hynes}},\ and\
  \bibinfo {author} {\bibfnamefont {R.}~\bibnamefont {Kapral}},\ }\bibfield
  {title} {\enquote {\bibinfo {title} {Dynamics of ion pair interconversion in
  a polar solvent},}\ }\href {https://doi.org/10.1063/1.459437} {\bibfield
  {journal} {\bibinfo  {journal} {J. Chem. Phys.}\ }\textbf {\bibinfo {volume}
  {93}},\ \bibinfo {pages} {7137--7147} (\bibinfo {year} {1990})}\BibitemShut
  {NoStop}%
\bibitem [{\citenamefont {Rey}\ and\ \citenamefont {Guardia}(1992)}]{Rey1992a}%
  \BibitemOpen
  \bibfield  {author} {\bibinfo {author} {\bibfnamefont {R.}~\bibnamefont
  {Rey}}\ and\ \bibinfo {author} {\bibfnamefont {E.}~\bibnamefont {Guardia}},\
  }\bibfield  {title} {\enquote {\bibinfo {title} {Dynamical aspects of the
  sodium(1$+$)-chloride ion pair association in water},}\ }\href
  {https://doi.org/10.1021/j100190a104} {\bibfield  {journal} {\bibinfo
  {journal} {J. Phys. Chem.}\ }\textbf {\bibinfo {volume} {96}},\ \bibinfo
  {pages} {4712--4718} (\bibinfo {year} {1992})}\BibitemShut {NoStop}%
\bibitem [{\citenamefont {Bergsma}\ \emph {et~al.}(1987)\citenamefont
  {Bergsma}, \citenamefont {Gertner}, \citenamefont {Wilson},\ and\
  \citenamefont {Hynes}}]{bergsma87}%
  \BibitemOpen
  \bibfield  {author} {\bibinfo {author} {\bibfnamefont {J.~P.}\ \bibnamefont
  {Bergsma}}, \bibinfo {author} {\bibfnamefont {B.~J.}\ \bibnamefont
  {Gertner}}, \bibinfo {author} {\bibfnamefont {K.~R.}\ \bibnamefont
  {Wilson}},\ and\ \bibinfo {author} {\bibfnamefont {J.~T.}\ \bibnamefont
  {Hynes}},\ }\bibfield  {title} {\enquote {\bibinfo {title} {Molecular
  dynamics of a model {S\textsubscript{N}2} reaction in water},}\ }\href
  {https://doi.org/10.1063/1.452224} {\bibfield  {journal} {\bibinfo  {journal}
  {J. Chem. Phys.}\ }\textbf {\bibinfo {volume} {86}},\ \bibinfo {pages} {1356}
  (\bibinfo {year} {1987})}\BibitemShut {NoStop}%
\bibitem [{\citenamefont {Gertner}\ \emph {et~al.}(1987)\citenamefont
  {Gertner}, \citenamefont {Bergsma}, \citenamefont {Wilson}, \citenamefont
  {Lee},\ and\ \citenamefont {Hynes}}]{hynes87}%
  \BibitemOpen
  \bibfield  {author} {\bibinfo {author} {\bibfnamefont {B.~J.}\ \bibnamefont
  {Gertner}}, \bibinfo {author} {\bibfnamefont {J.~P.}\ \bibnamefont
  {Bergsma}}, \bibinfo {author} {\bibfnamefont {K.~R.}\ \bibnamefont {Wilson}},
  \bibinfo {author} {\bibfnamefont {S.}~\bibnamefont {Lee}},\ and\ \bibinfo
  {author} {\bibfnamefont {J.~T.}\ \bibnamefont {Hynes}},\ }\bibfield  {title}
  {\enquote {\bibinfo {title} {Molecular dynamics of a model
  {S\textsubscript{N}2} reaction in water},}\ }\href
  {https://doi.org/10.1063/1.452224} {\bibfield  {journal} {\bibinfo  {journal}
  {J. Chem. Phys.}\ }\textbf {\bibinfo {volume} {86}},\ \bibinfo {pages} {1377}
  (\bibinfo {year} {1987})}\BibitemShut {NoStop}%
\bibitem [{\citenamefont {Gertner}, \citenamefont {Wilson},\ and\ \citenamefont
  {Hynes}(1989)}]{Gertner1989a}%
  \BibitemOpen
  \bibfield  {author} {\bibinfo {author} {\bibfnamefont {B.~J.}\ \bibnamefont
  {Gertner}}, \bibinfo {author} {\bibfnamefont {K.~R.}\ \bibnamefont
  {Wilson}},\ and\ \bibinfo {author} {\bibfnamefont {J.~T.}\ \bibnamefont
  {Hynes}},\ }\bibfield  {title} {\enquote {\bibinfo {title} {Nonequilibrium
  solvation effects on reaction rates for model {S\textsubscript{N}2} reactions
  in water},}\ }\href@noop {} {\bibfield  {journal} {\bibinfo  {journal} {J.
  Chem. Phys.}\ }\textbf {\bibinfo {volume} {90}},\ \bibinfo {pages}
  {3537--3558} (\bibinfo {year} {1989})},\ \Eprint
  {https://arxiv.org/abs/doi10.1063/1.455864} {doi10.1063/1.455864}
  \BibitemShut {NoStop}%
\bibitem [{\citenamefont {Roca}\ \emph {et~al.}(2006)\citenamefont {Roca},
  \citenamefont {Moliner}, \citenamefont {Tu{\~{n}}{\'{o}}n},\ and\
  \citenamefont {Hynes}}]{Roca2006a}%
  \BibitemOpen
  \bibfield  {author} {\bibinfo {author} {\bibfnamefont {M.}~\bibnamefont
  {Roca}}, \bibinfo {author} {\bibfnamefont {V.}~\bibnamefont {Moliner}},
  \bibinfo {author} {\bibfnamefont {I.}~\bibnamefont {Tu{\~{n}}{\'{o}}n}},\
  and\ \bibinfo {author} {\bibfnamefont {J.~T.}\ \bibnamefont {Hynes}},\
  }\bibfield  {title} {\enquote {\bibinfo {title} {Coupling between protein and
  reaction dynamics in enzymatic processes: Application of {Grote--Hynes}
  theory to catechol {O}-methyltransferase},}\ }\href
  {https://doi.org/10.1021/ja058826u} {\bibfield  {journal} {\bibinfo
  {journal} {J. Am. Chem. Soc.}\ }\textbf {\bibinfo {volume} {128}},\ \bibinfo
  {pages} {6186--6193} (\bibinfo {year} {2006})}\BibitemShut {NoStop}%
\bibitem [{\citenamefont {Ruiz-Pern{\'{\i}}a}\ \emph
  {et~al.}(2008)\citenamefont {Ruiz-Pern{\'{\i}}a}, \citenamefont
  {Tu{\~{n}}{\'{o}}n}, \citenamefont {Moliner}, \citenamefont {Hynes},\ and\
  \citenamefont {Roca}}]{RuizPernia2008a}%
  \BibitemOpen
  \bibfield  {author} {\bibinfo {author} {\bibfnamefont {J.~J.}\ \bibnamefont
  {Ruiz-Pern{\'{\i}}a}}, \bibinfo {author} {\bibfnamefont {I.}~\bibnamefont
  {Tu{\~{n}}{\'{o}}n}}, \bibinfo {author} {\bibfnamefont {V.}~\bibnamefont
  {Moliner}}, \bibinfo {author} {\bibfnamefont {J.~T.}\ \bibnamefont {Hynes}},\
  and\ \bibinfo {author} {\bibfnamefont {M.}~\bibnamefont {Roca}},\ }\bibfield
  {title} {\enquote {\bibinfo {title} {Dynamic effects on reaction rates in a
  michael addition catalyzed by chalcone isomerase. beyond the frozen
  environment approach},}\ }\href {https://doi.org/10.1021/ja801156y}
  {\bibfield  {journal} {\bibinfo  {journal} {J. Am. Chem. Soc.}\ }\textbf
  {\bibinfo {volume} {130}},\ \bibinfo {pages} {7477--7488} (\bibinfo {year}
  {2008})}\BibitemShut {NoStop}%
\bibitem [{\citenamefont {Kanaan}\ \emph {et~al.}(2010)\citenamefont {Kanaan},
  \citenamefont {Roca}, \citenamefont {Tu{\~{n}}{\'{o}}n}, \citenamefont
  {Mart{\'{\i}}},\ and\ \citenamefont {Moliner}}]{Kanaan2010a}%
  \BibitemOpen
  \bibfield  {author} {\bibinfo {author} {\bibfnamefont {N.}~\bibnamefont
  {Kanaan}}, \bibinfo {author} {\bibfnamefont {M.}~\bibnamefont {Roca}},
  \bibinfo {author} {\bibfnamefont {I.}~\bibnamefont {Tu{\~{n}}{\'{o}}n}},
  \bibinfo {author} {\bibfnamefont {S.}~\bibnamefont {Mart{\'{\i}}}},\ and\
  \bibinfo {author} {\bibfnamefont {V.}~\bibnamefont {Moliner}},\ }\bibfield
  {title} {\enquote {\bibinfo {title} {Application of {Grote--Hynes} theory to
  the reaction catalyzed by thymidylate synthase},}\ }\href
  {https://doi.org/10.1021/jp1072457} {\bibfield  {journal} {\bibinfo
  {journal} {J. Phys. Chem. B}\ }\textbf {\bibinfo {volume} {114}},\ \bibinfo
  {pages} {13593--13600} (\bibinfo {year} {2010})}\BibitemShut {NoStop}%
\bibitem [{\citenamefont {Fukui}(1970)}]{Fukui1970}%
  \BibitemOpen
  \bibfield  {author} {\bibinfo {author} {\bibfnamefont {K.}~\bibnamefont
  {Fukui}},\ }\bibfield  {title} {\enquote {\bibinfo {title} {Formulation of
  the reaction coordinate},}\ }\href {https://doi.org/10.1021/j100717a029}
  {\bibfield  {journal} {\bibinfo  {journal} {J. Phys. Chem.}\ }\textbf
  {\bibinfo {volume} {74}},\ \bibinfo {pages} {4161--4163} (\bibinfo {year}
  {1970})}\BibitemShut {NoStop}%
\bibitem [{\citenamefont {Nagahata}\ \emph {et~al.}(2020)\citenamefont
  {Nagahata}, \citenamefont {Borondo}, \citenamefont {Benito},\ and\
  \citenamefont {Hernandez}}]{hern20f}%
  \BibitemOpen
  \bibfield  {author} {\bibinfo {author} {\bibfnamefont {Y.}~\bibnamefont
  {Nagahata}}, \bibinfo {author} {\bibfnamefont {F.}~\bibnamefont {Borondo}},
  \bibinfo {author} {\bibfnamefont {R.~M.}\ \bibnamefont {Benito}},\ and\
  \bibinfo {author} {\bibfnamefont {R.}~\bibnamefont {Hernandez}},\ }\bibfield
  {title} {\enquote {\bibinfo {title} {Identifying reaction pathways via
  asymptotic trajectories},}\ }\href {https://doi.org/10.1039/c9cp06610a}
  {\bibfield  {journal} {\bibinfo  {journal} {Phys. Chem. Chem. Phys.}\
  }\textbf {\bibinfo {volume} {22}},\ \bibinfo {pages} {10087--10105} (\bibinfo
  {year} {2020})}\BibitemShut {NoStop}%
\bibitem [{\citenamefont {Wormer}\ and\ \citenamefont
  {Tennyson}(1981)}]{wormer1981abinitio}%
  \BibitemOpen
  \bibfield  {author} {\bibinfo {author} {\bibfnamefont {P.~E.~S.}\
  \bibnamefont {Wormer}}\ and\ \bibinfo {author} {\bibfnamefont
  {J.}~\bibnamefont {Tennyson}},\ }\bibfield  {title} {\enquote {\bibinfo
  {title} {Ab initio {SCF} calculations on the potential energy surface of
  potassium cyanide ({KCN})},}\ }\href {https://doi.org/10.1063/1.442174}
  {\bibfield  {journal} {\bibinfo  {journal} {J. Chem. Phys.}\ }\textbf
  {\bibinfo {volume} {75}},\ \bibinfo {pages} {1245--1252} (\bibinfo {year}
  {1981})}\BibitemShut {NoStop}%
\bibitem [{\citenamefont {Tennyson}()}]{TennysonLiCNProcedure}%
  \BibitemOpen
  \bibfield  {author} {\bibinfo {author} {\bibfnamefont {J.}~\bibnamefont
  {Tennyson}},\ }\href@noop {} {\enquote {\bibinfo {title} {{LiCN/LiNC}
  potential surface procedure},}\ }\bibinfo {note} {Private
  communication}\BibitemShut {NoStop}%
\bibitem [{\citenamefont {Reiff}(2021)}]{reiff21licn}%
  \BibitemOpen
  \bibfield  {author} {\bibinfo {author} {\bibfnamefont {J.}~\bibnamefont
  {Reiff}},\ }\href {https://github.com/joreiff/licn-potential} {\enquote
  {\bibinfo {title} {{LiCN/LiNC} isomerization potential surface},}\ }
  (\bibinfo {year} {2021})\BibitemShut {NoStop}%
\bibitem [{\citenamefont {Arrhenius}(1889)}]{Arrhenius1889}%
  \BibitemOpen
  \bibfield  {author} {\bibinfo {author} {\bibfnamefont {S.}~\bibnamefont
  {Arrhenius}},\ }\bibfield  {title} {\enquote {\bibinfo {title} {{\"U}ber die
  {R}eaktionsgeschwindigkeit bei der {I}nversion von {R}ohzucker durch
  {S\"a}uren},}\ }\href {https://doi.org/10.1016/B978-0-08-012344-8.50005-2}
  {\bibfield  {journal} {\bibinfo  {journal} {Z. Phys. Chem.}\ }\textbf
  {\bibinfo {volume} {4U}},\ \bibinfo {pages} {226--248} (\bibinfo {year}
  {1889})},\ \bibinfo {note} {translated and published in Margaret H. Back and
  Keith J. Laidler, eds., Selected Readings in Chemical Kinetics (Oxford:
  Pergamon, 1967)}\BibitemShut {NoStop}%
\bibitem [{\citenamefont {Polanyi}\ and\ \citenamefont
  {Wigner}(1928)}]{Polanyi1928a}%
  \BibitemOpen
  \bibfield  {author} {\bibinfo {author} {\bibfnamefont {M.}~\bibnamefont
  {Polanyi}}\ and\ \bibinfo {author} {\bibfnamefont {E.}~\bibnamefont
  {Wigner}},\ }\bibfield  {title} {\enquote {\bibinfo {title} {{\"Uber die
  Interferenz von Eigenschwingungen als Ursache von Energieschwankungen und
  chemischer Umsetzungen}},}\ }\href {https://doi.org/10.1515/zpch-1928-13930}
  {\bibfield  {journal} {\bibinfo  {journal} {Z. Phys. Chem.}\ }\textbf
  {\bibinfo {volume} {139A}},\ \bibinfo {pages} {439--452} (\bibinfo {year}
  {1928})}\BibitemShut {NoStop}%
\bibitem [{\citenamefont {Eyring}(1935{\natexlab{b}})}]{eyring35b}%
  \BibitemOpen
  \bibfield  {author} {\bibinfo {author} {\bibfnamefont {H.}~\bibnamefont
  {Eyring}},\ }\bibfield  {title} {\enquote {\bibinfo {title} {The activated
  complex and the absolute rate of chemical reactions},}\ }\href
  {https://doi.org/10.1021/cr60056a006} {\bibfield  {journal} {\bibinfo
  {journal} {Chem. Rev.}\ }\textbf {\bibinfo {volume} {17}},\ \bibinfo {pages}
  {65--77} (\bibinfo {year} {1935}{\natexlab{b}})}\BibitemShut {NoStop}%
\bibitem [{\citenamefont {{K. J. Laidler}}\ and\ \citenamefont {{M. C.
  King}}(1983)}]{Laidler1983}%
  \BibitemOpen
  \bibfield  {author} {\bibinfo {author} {\bibnamefont {{K. J. Laidler}}}\ and\
  \bibinfo {author} {\bibnamefont {{M. C. King}}},\ }\bibfield  {title}
  {\enquote {\bibinfo {title} {The development of transition-state theory},}\
  }\href {https://doi.org/10.1021/j100238a002} {\bibfield  {journal} {\bibinfo
  {journal} {J. Phys. Chem.}\ }\textbf {\bibinfo {volume} {87}},\ \bibinfo
  {pages} {2657--2664} (\bibinfo {year} {1983})}\BibitemShut {NoStop}%
\end{thebibliography}%

\end{document}